\documentclass[fleqn,usenatbib]{mnras} 

\usepackage{epstopdf, graphicx, color}
\usepackage{amssymb}
\usepackage{amsfonts}
\usepackage{amsbsy}
\usepackage{amsmath}
\usepackage{tabularx}
\usepackage{booktabs}
\usepackage{lineno}
\usepackage{eso-pic}

\usepackage{newtxtext,newtxmath}
\usepackage[T1]{fontenc}
\usepackage{ae,aecompl}

\newcommand{\nord}[1]{{\color{black} #1}}

\def \diehlpaper {{\it Diehl17}}

\def \nbcandidate		{46}
\def \nbfollowup 		{21}

\def \nbuncertain       {10}

\def \nbconfirmtext     {nine}
\def \nbrejecttext      {two}
\def \nbuncertaintext   {ten}

\def \figurescalemultipanel {1.0}
\def \figurescalesystem {0.8}

\def \Hbeta		{H$\beta$}
\def \Hgamma 	{H$\gamma$}
\def \Hdelta 	{H$\delta$}
\def \Hbeta 	{H$\beta$}
\def \Lya 		{Ly$\alpha$}
\def \OII 		{$[{\rm OII}]$3727}
\def \OIIdoublet{[OII]}
\def \OIIIa 	{[OIII]4959}
\def \OIIIb 	{[OIII]5006}
\def \OIII		{[OIII]}
\def \NeIII 	{[NeIII]3869}
\def \NeIV 		{[NeIV]2424}
\def \NeV 		{[NeV]3346}

\def \CII 		{CII]2326}

\def \zlens       {z_{\rm lens}}
\def \zsrc     	  {z_{\rm source}}
\def \zspec 	  {z_{\rm spec}}
\def \thetae      {\theta_{\rm e}}
\def \thetaesep   {\theta_{\rm sep}}
\def \menclosed   {M_{\rm enc}}

\def \DL 		  {D_{\rm L}}
\def \DS 		  {D_{\rm S}}
\def \DLS 		  {D_{\rm LS}}

\def \redmagic 	  {redMaGiC}
\def \redmapper   {redMaPPer}

\def \einsteinradiussub {source image-lens separation}

\def \kms  {{\rm km\, s^{-1}}}
\def \msol {{\rm M_{\odot}}}
\def \sqdeg {sq. deg.} 
\def \sqdegadjective {-sq.-deg.\ }
\def \sqarcsec {{\rm arcsec}^2}

\def \arcsec {''}

% ... 
\def \FigureSystemDirectory {FigureSystems} 
\def \FigurePanelDirectory {FigurePanel} 

\def \arcA {A1}
\def \arcB {A2}
\def \arcC {A3}

\def \DESSystemName {DES System Name}
\def \ObjectName {Object Name}

\newcommand\T{\rule{0pt}{3ex}}       
\newcommand\M{\rule{0pt}{3ex}}       

\def \SystemA {\mbox{DES J0041-4155}}
\def \SystemC {\mbox{DES J0120-5143}}
\def \SystemB {\mbox{DES J0104-5341}}
\def \SystemE {\mbox{DES J2113-0114}}
\def \SystemD {\mbox{DES J0357-4756}}
\def \SystemG {\mbox{DES J2349-5113}}
\def \SystemF {\mbox{DES J2321-4630}}
\def \SystemH {\mbox{DES J0418-5457}}
\def \SystemI {\mbox{DES J0227-4516}}

\def \SystemARadius {7.23 \pm 0.50}
\def \SystemBRadius {2.18 \pm 0.43}
\def \SystemCRadius {3.35 \pm 0.51}
\def \SystemDRadius {9.39 \pm 0.91}
\def \SystemERadius {1.98 \pm 0.51}
\def \SystemFRadius {3.30 \pm 0.74}
\def \SystemGRadius {4.46 \pm 0.71}
\def \SystemHRadius {1.97 \pm 0.27}
\def \SystemIRadius {4.11 \pm 0.28}

\def \SystemAMass {1.74 \pm 0.47 \times 10^{13}\, \msol}
\def \SystemBMass {2.42 \pm 1.10 \times 10^{12}\, \msol}
\def \SystemCMass {4.46 \pm 2.78 \times 10^{12}\, \msol}
\def \SystemDMass {1.38 \pm 0.36 \times 10^{13}\, \msol}
\def \SystemEMass {1.28 \pm 1.06 \times 10^{12}\, \msol}
\def \SystemFMass {3.84 \pm 1.87 \times 10^{12}\, \msol}
\def \SystemGMass {3.77 \pm 1.42 \times 10^{12}\, \msol}
\def \SystemHMass {1.47 \pm 0.65 \times 10^{12}\, \msol}
\def \SystemIMass {4.32 \pm 1.32 \times 10^{12}\, \msol}

\def \redshiftlensA {0.7160 \pm 0.0310}
\def \redshiftsrcAA {2.5619 \pm 0.0001}
\def \redshiftsrcAB {2.5618 \pm 0.0001}
\def \redshiftsrcAC {2.5616 \pm 0.0002}

\def \redshiftlensB {0.6790 \pm 0.0220}
\def \redshiftsrcBA {1.2318 \pm 0.0001}
\def \redshiftsrcBB {1.2318 \pm 0.0001}

\def \redshiftlensC {0.6030 \pm 0.0630}
\def \redshiftsrcCA {1.2955 \pm 0.0001}
\def \redshiftsrcCB {1.2957 \pm 0.0000}
\def \redshiftsrcCC {1.2957 \pm 0.0005}

\def \redshiftlensD {0.2570 \pm 0.0240}
\def \redshiftsrcDA {0.9156 \pm 0.0001}
\def \redshiftsrcDB {0.9156 \pm 0.0001}
\def \redshiftsrcDC {0.9155 \pm 0.0001}

\def \redshiftlensE {0.4060 \pm 0.0710}
\def \redshiftsrcEA {0.7943 \pm 0.0002}
\def \redshiftsrcEB {0.7947 \pm 0.0002}

\def \redshiftlensFA {0.6440 \pm 0.0250}
\def \redshiftlensFB {0.70\pm0.02}
\def \redshiftsrcFA {1.7469 \pm 0.0014}
\def \redshiftsrcFB {1.7479 \pm 0.0031}

\def \redshiftlensG {0.3450 \pm 0.0320}
\def \redshiftsrcGA {1.3932 \pm 0.0001}
\def \redshiftsrcGB {1.3938 \pm 0.0001}

\def \redshiftlensH {0.6130 \pm 0.0430}
\def \redshiftsrcHA {1.4236 \pm 0.0001}

\def \redshiftlensI {0.4300 \pm 0.0400}
\def \redshiftsrcIA {1.25264 \pm 0.00009}

\def \redshiftRMA {0.7317 \pm 0.0198} 
\def \redshiftRMB {0.6500 \pm 0.0259} 
\def \redshiftRMC {0.5238 \pm 0.0173} 
\def \redshiftRMD {0.2755 \pm 0.0129} 
\def \redshiftRME {0.4472 \pm 0.0187} 
\def \redshiftRMF {0.6427 \pm 0.0363} 
\def \redshiftRMG {0.4094 \pm 0.0189} 
\def \redshiftRMH {\redshiftlensH} 
\def \redshiftRMI {0.4347 \pm 0.0183} 

\def \redshiftRMFa {0.6244 \pm 0.0240} 

\def \figuremultipanel 
{
\begin{figure*}
\centering
\includegraphics[scale=\figurescalemultipanel]{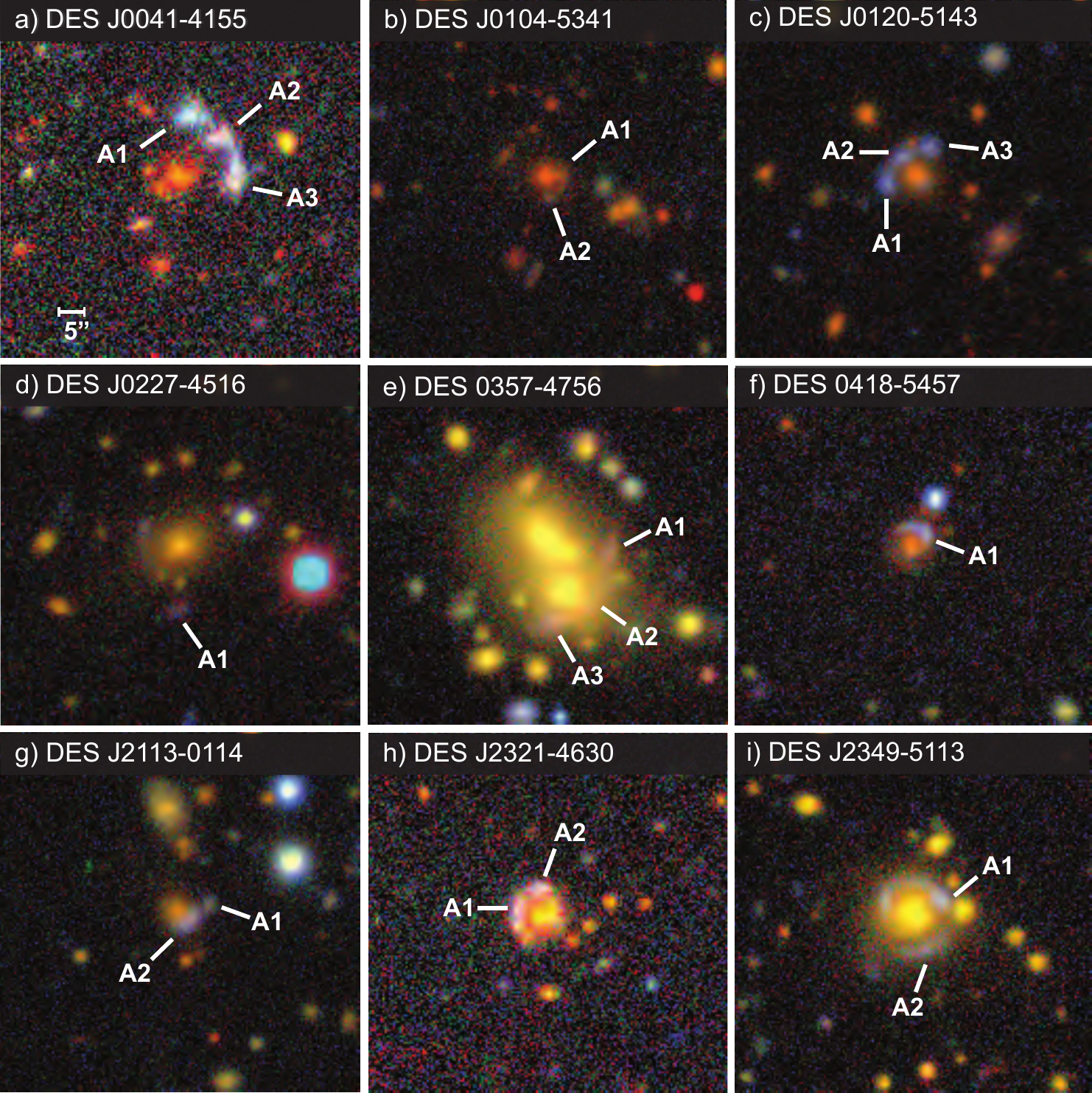}
\caption{Color co-added DES images of the \nbconfirmtext\ confirmed lensing systems described in this work: (a)\SystemA, (b)\SystemB, (c)\SystemC, (d)\SystemI, (e)\SystemD, (f)\SystemH, (g)\SystemE, (h)\SystemF, (i)\SystemG. Images are oriented north up, east left, and are $1'\times 1'$ in area. The color images are composites of {\it g}, {\it r}, and {\it i} bands. The lensing features targeted for spectroscopy are indicated by the letters. Refer to Fig.'s~\ref{fig:SystemA}-\ref{fig:SystemG} for additional images of the candidates and their spectroscopic information.}
\label{fig:multipanel}
\end{figure*}
}

\def \FigureSystemA
{
\begin{figure*}
\centering
\includegraphics[scale=\figurescalesystem]{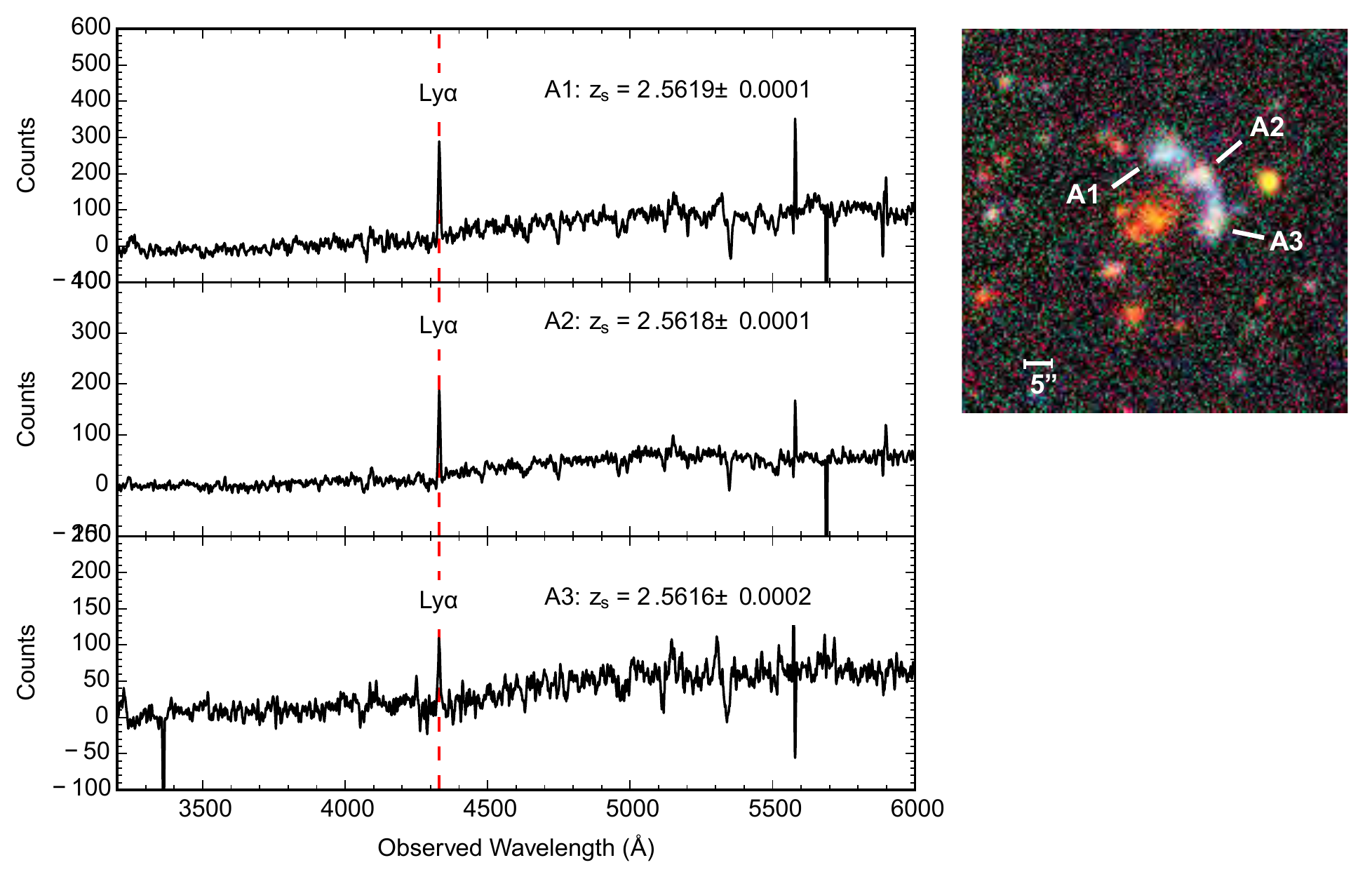}
\caption{\SystemA. The 1D spectra for \arcA, \arcB, and \arcC\ are shown on the left. There are emission lines near the same observed wavelength $\sim4329$\AA\ in all three B600 spectra. We assign these features to be \Lya, which gives redshifts of $\zsrc=\redshiftsrcAA$, $\redshiftsrcAB$, and $\redshiftsrcAC$ for \arcA, \arcB, and \arcC, respectively. The spectra are smoothed using a boxcar with a width of 5 pixels with the IRAF task {\tt splot}. In the color coadd image in the top right panel, the three features of interest are labeled, and the scale bar shows the size of the image. In the lower right is an identical color coadd image. Both images are oriented north up, east left.   }
\label{fig:SystemA}
\end{figure*}
}

\def \FigureSystemB
{
\begin{figure*}
\centering
\includegraphics[scale=\figurescalesystem]{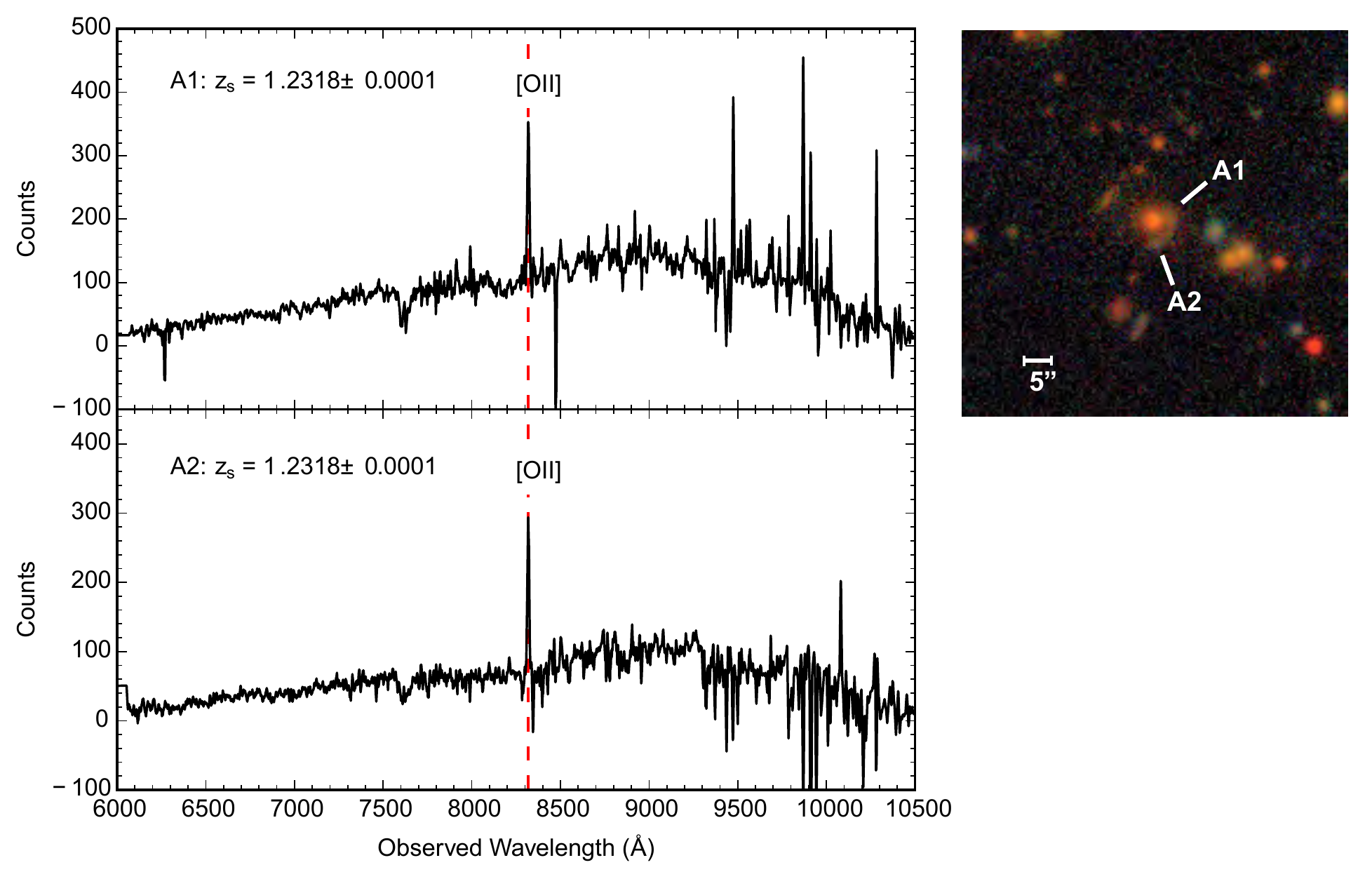}
\caption{\SystemB. The 1D spectra for source images \arcA\ and \arcB\ are shown on the left. There are emission lines in both R400 spectra near an observed wavelength of $\sim8319$\AA. We associate this feature with \OIIdoublet, which yields redshifts of $\zsrc=\redshiftsrcBA$ and $\redshiftsrcBA$ in \arcA\ and \arcB, respectively. The spectra are smoothed using a boxcar with a width of 5 pixels with the IRAF task {\tt splot}.In the color coadd image in the top right panel, the two features of interest are labeled ``\arcA'' and ``\arcB'', and the scale bar shows the size of the image. In the lower right is an identical color coadd image. Both images are oriented north up, east left. }
\label{fig:SystemB}
\end{figure*}
}

\def \FigureSystemC
{
\begin{figure*}
\centering
\includegraphics[scale=\figurescalesystem]{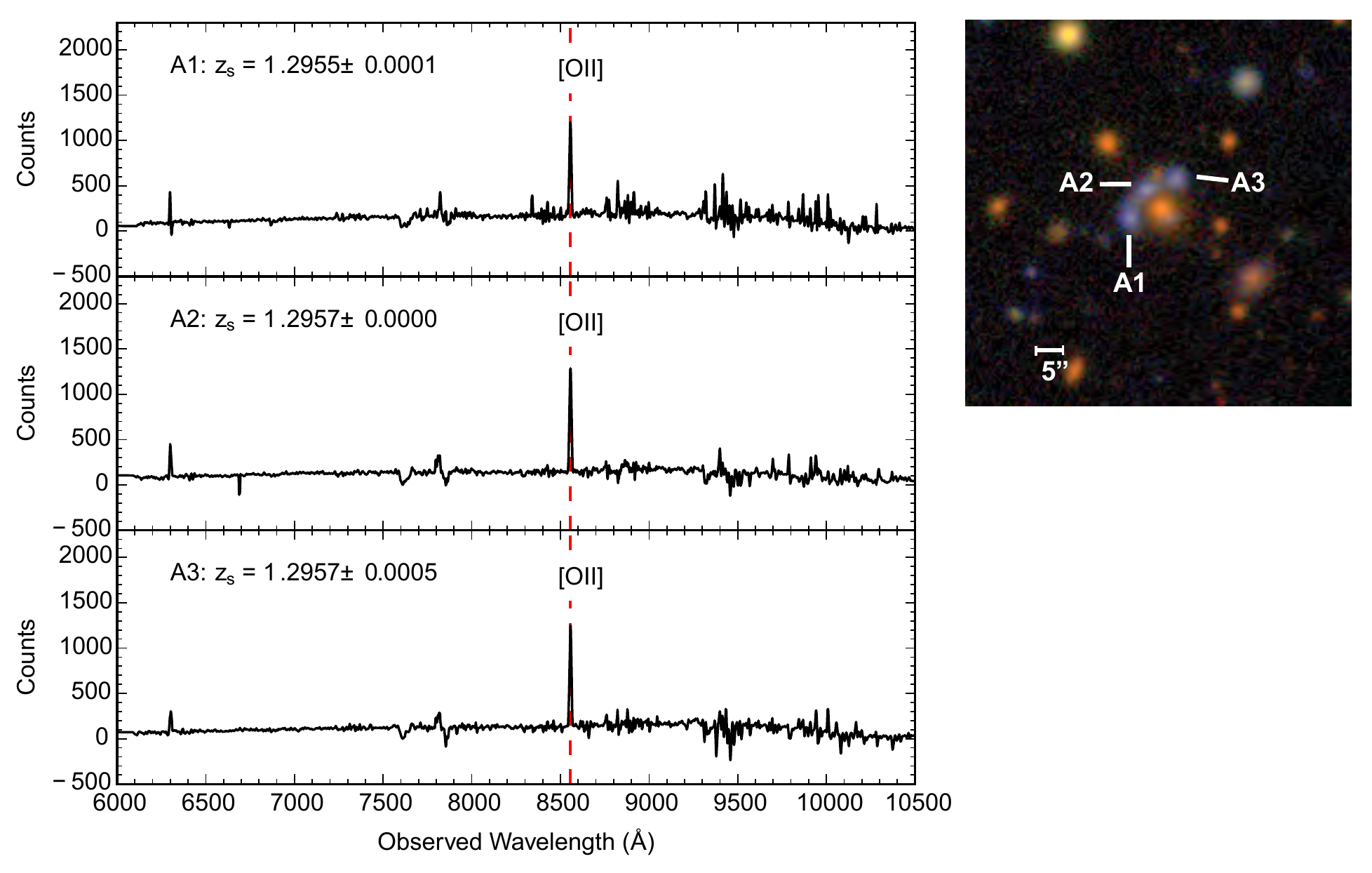}
\caption{\SystemC. The 1D spectra for source images \arcA, \arcB, and \arcC\ are shown on the left. There are emission lines in all three R400 spectra near $\sim8556$\AA, which we identify as \OIIdoublet, and from which we obtain redshifts, $\zsrc=\redshiftsrcCA$, $\redshiftsrcCB$, and $\redshiftsrcCC$  for \arcA, \arcB, and \arcC, respectively. The spectra are smoothed using a boxcar with a width of 5 pixels with the IRAF task {\tt splot}. In the color coadd image in the top right panel, the three features of interest are labeled ``\arcA,'' ``\arcB,'' and ``\arcC.'' The scale bar shows the size of the image. The image is oriented north up, east left. }
\label{fig:SystemC}
\end{figure*}
}

\def \FigureSystemD 
{
\begin{figure*}
\centering
\includegraphics[scale=\figurescalesystem]{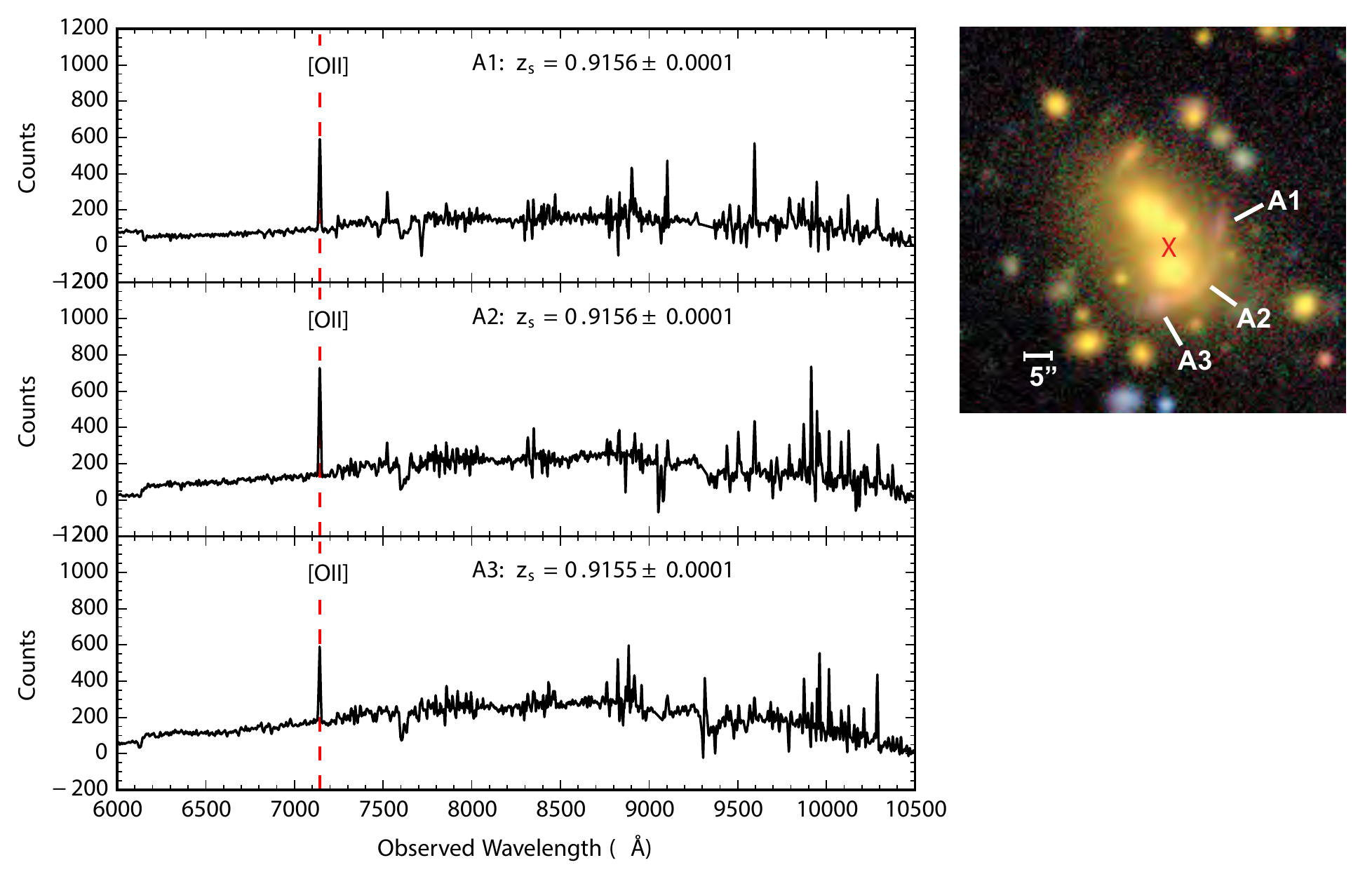}
\caption{\SystemD. The 1D spectra for source images \arcA, \arcB, and \arcC\ are shown on the left. Near the the observed wavelength of $\sim7142$\AA\ we identify \OIIdoublet\ in all three R400 spectra, from which we obtain redshifts of $\zsrc=\redshiftsrcDA$, $\redshiftsrcDB$, and $\redshiftsrcDC$, for \arcA, \arcB, and \arcC, respectively, and which are labeled in the color coadd image in the top right panel, along with a scale bar to show the size of the image. The spectra are smoothed using a boxcar with a width of 5 pixels with the IRAF task {\tt splot}. 
The image is oriented north up, east left. The red 'x' near the center of the color image in the upper right panel marks the position designated for measurement of the Einstein radius. 
The position of the red ‘x’ is chosen to simplify the drawing of a circle through the arcs of the source galaxy images to extract the Einstein radius and lens mass.}
\label{fig:SystemD}
\end{figure*}
}

\def \FiguresystemE
{
\begin{figure*}
	\centering
	\includegraphics[scale=\figurescalesystem]{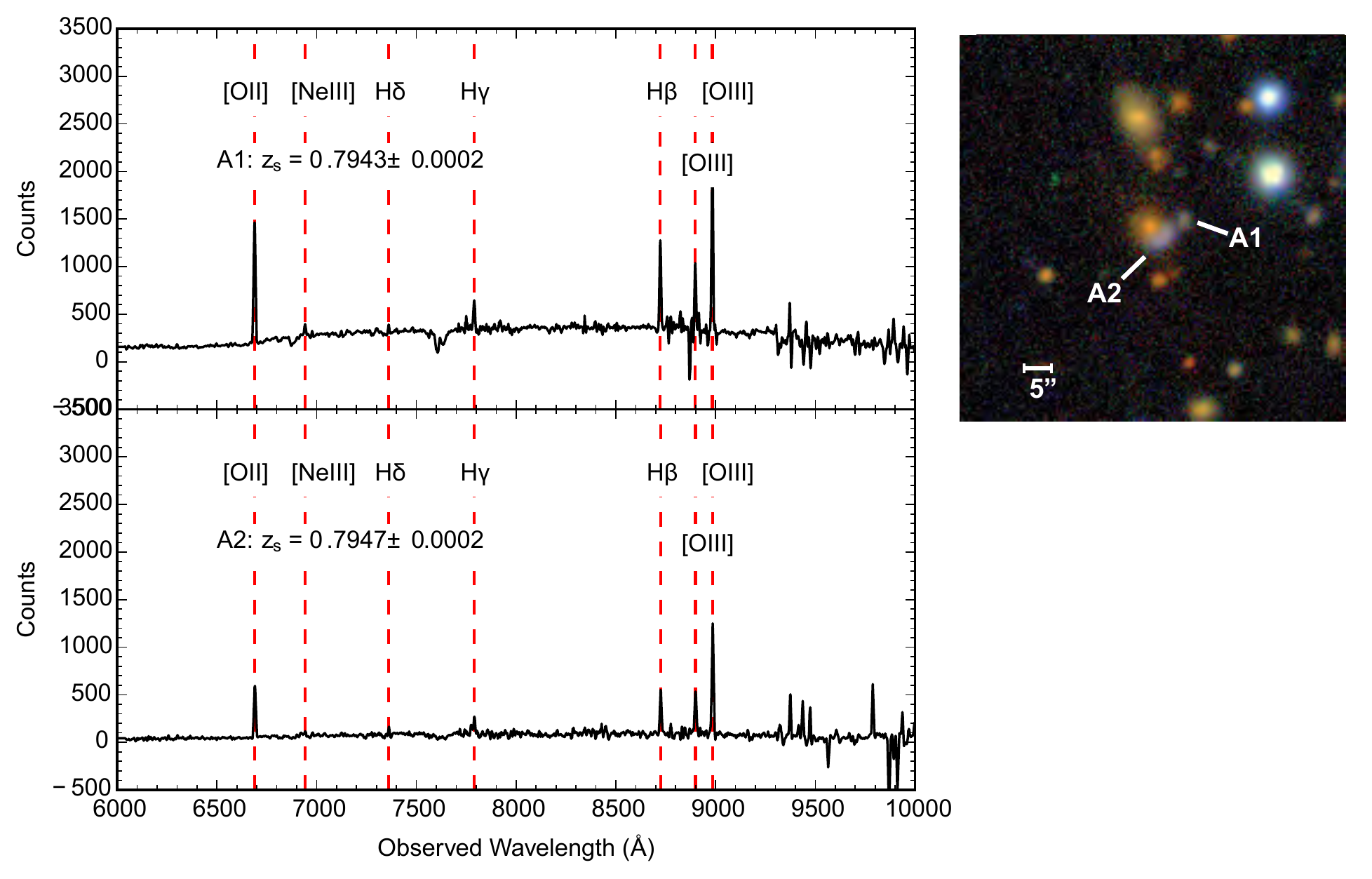}
	\caption{\SystemE. The 1D R400 spectra for source images \arcA\ and \arcB\ are shown on the left. \arcA\ presents emission lines near observed wavelengths of $\sim6689$\AA, $\sim6942$\AA, $\sim7790$\AA, $\sim8723$\AA, $\sim8898$\AA, $\sim8984$\AA, which we conclude correspond to \OIIdoublet, \NeIII, \Hdelta, \Hgamma, \Hbeta, \OIIIa, and \OIIIb. From these emission lines we obtain redshifts of $\zsrc=\redshiftsrcEA$ for \arcA\ and $\redshiftsrcEB$ for \arcB. The spectra are smoothed using a boxcar with a width of 5 pixels with the IRAF task {\tt splot}. In the color coadd image in the top right panel the two features of interest are labeled ``\arcA'' and ``\arcB,'' along with a scale bar to show the size of the image. The image is oriented north up, east left. }
	\label{fig:SystemE}
\end{figure*}
}

\def \FigureSystemF
{
\begin{figure*}
\centering
\includegraphics[scale=\figurescalesystem]{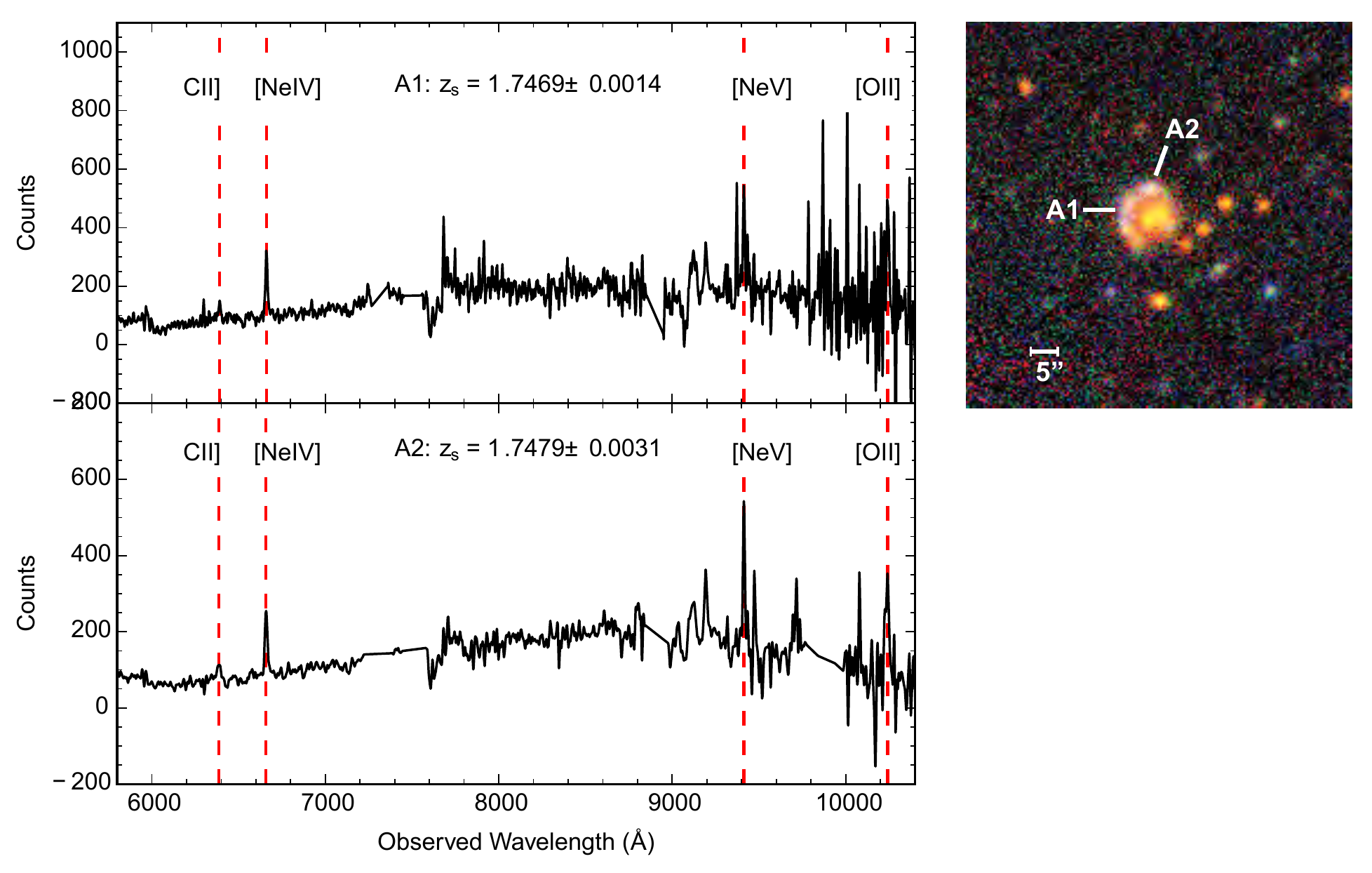}
\caption{\SystemF. The 1D R400 spectra for sources \arcA\ and \arcB\ are shown on the left. The emission lines at four different observed wavelengths in \arcA\ at $\sim6385$\AA, $\sim6658$\AA, $\sim9413$\AA, $\sim10242$\AA\ are due to \CII, \NeIV, and \OIIdoublet. A similar pattern of emission lines occurs in \arcB: $\sim6391$\AA, $\sim6661$\AA, $\sim9413$\AA, $\sim10242$\AA. These lines yield source spectroscopic redshifts of $\zsrc=\redshiftsrcFA$ and $\redshiftsrcFB$ for \arcA\ and \arcB, respectively. The spectra are smoothed using a boxcar with a width of 5 pixels with the IRAF task {\tt splot}. In the color coadd image in the top right panel the two features of interest are labeled ``\arcA'' and ``\arcB.'' The scale bar shows the size of the image. The image is oriented north up, east left. }
\label{fig:SystemF}
\end{figure*}
}

\def \FigureSystemG
{
\begin{figure*}
\centering
\includegraphics[scale=\figurescalesystem]{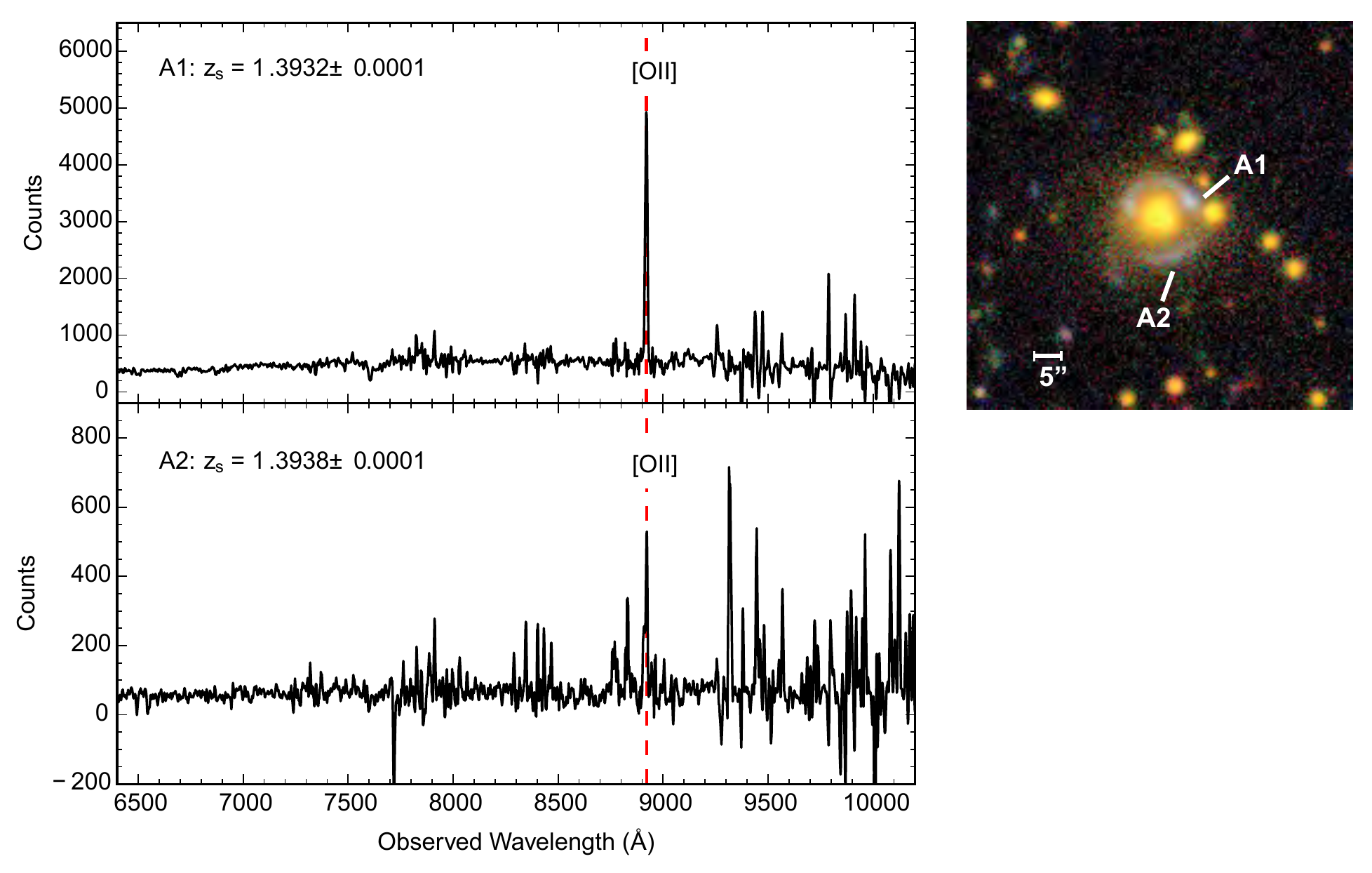}
\caption{\SystemG. The 1D R400 spectra for \arcA\ and \arcB\ are shown on the left. The prominent emission lines near observed wavelengths $\sim8920$\AA\  are considered to be \OIIdoublet. These lines yield source spectroscopic redshifts of $\zsrc=\redshiftsrcGA$ and $\redshiftsrcGB$ for \arcA\ and \arcB, respectively. The spectra are smoothed using a boxcar with a width of 5 pixels with the IRAF task {\tt splot}. In the color coadd image in the top right panel the two features of interest are labeled ``\arcA'' and ``\arcB.'' The scale bar shows the size of the image. The image is oriented north up, east left. }
\label{fig:SystemG}
\end{figure*}
}

\def \FigureSystemH
{
\begin{figure*}
\centering
\includegraphics[scale=\figurescalesystem]{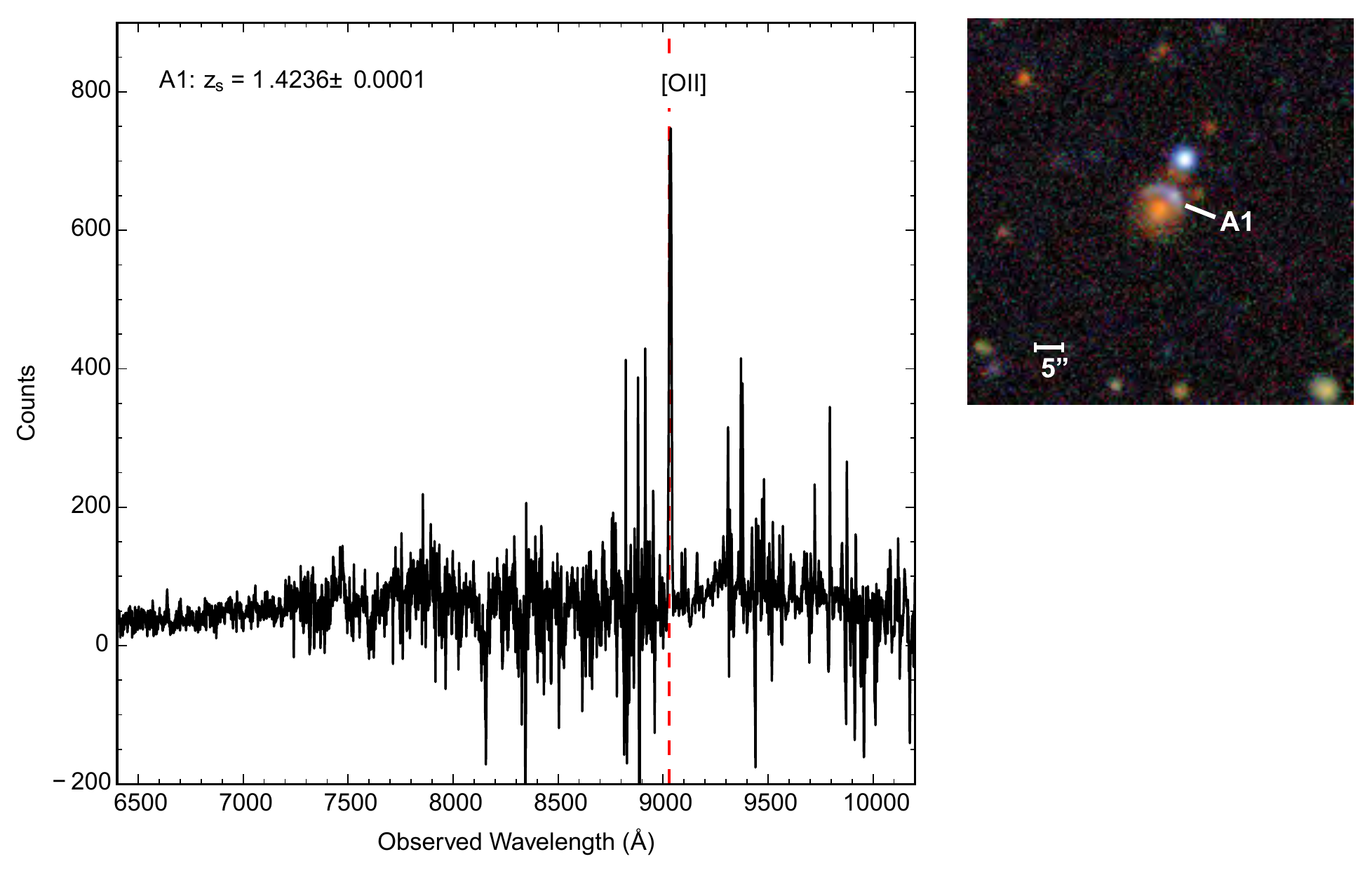}
\caption{\SystemH. The 1D R400 spectrum for \arcA\ is shown on the left. The prominent emission lines near observed wavelengths $\sim8920$\AA\  are considered to be \OIIdoublet. These lines yield source spectroscopic redshifts of $\zsrc=\redshiftsrcHA$ for \arcA. The spectra are smoothed using a boxcar with a width of 5 pixels with the IRAF task {\tt splot}. In the color coadd image in the top right panel, the feature of interest is labeled ``\arcA.'' The scale bar shows the size of the image. The image is oriented north up, east left. }
\label{fig:SystemH}
\end{figure*}
}

\def \FigureSystemI
{
\begin{figure*}
\centering
\includegraphics[scale=\figurescalesystem]{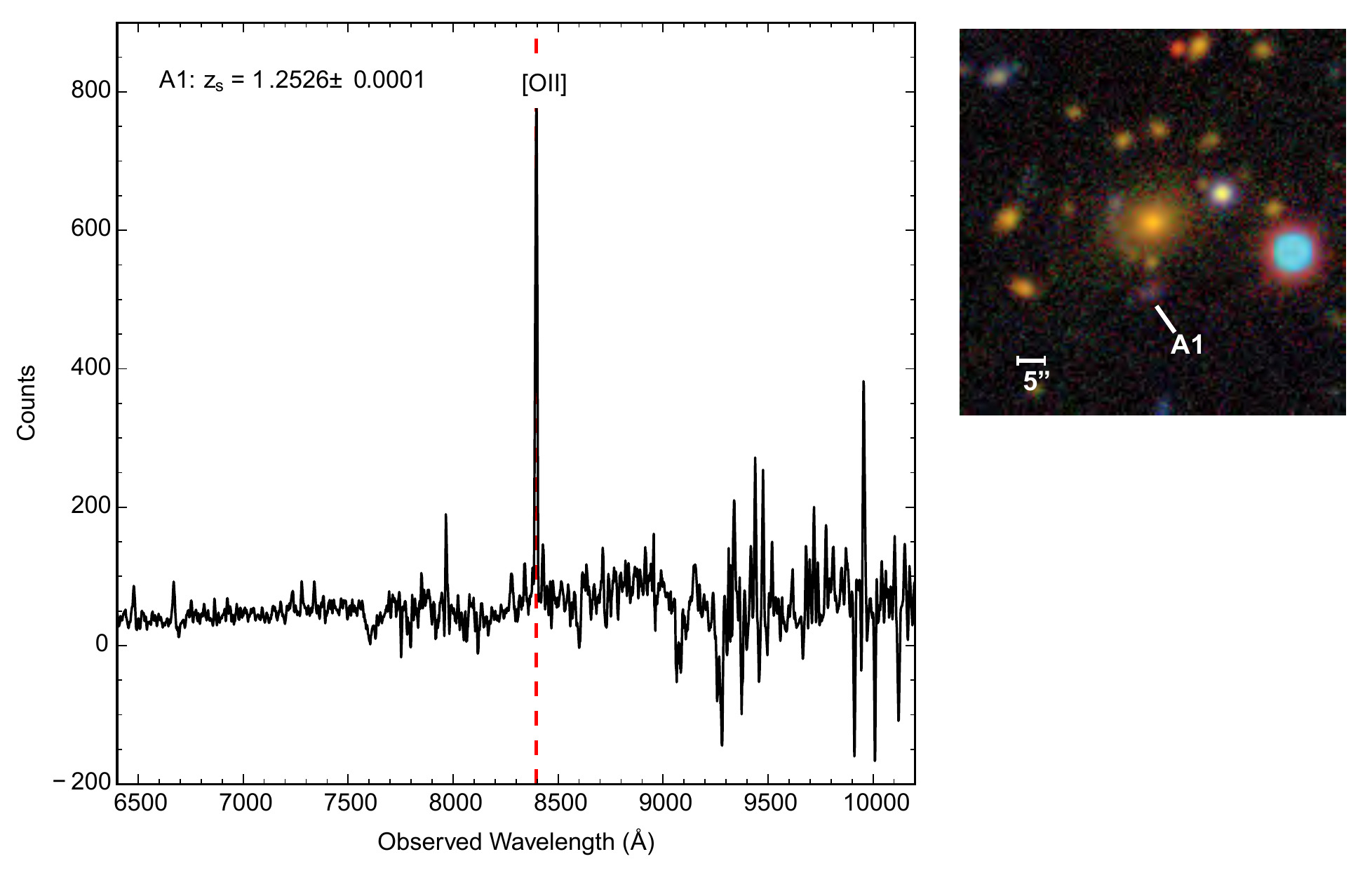}
\caption{\SystemI. The 1D R400 spectrum for \arcA\ is shown on the left. The prominent emission lines near observed wavelengths $\sim8396$\AA\  are considered to be \OIIdoublet. These lines yield source spectroscopic redshifts of $\zsrc=\redshiftsrcIA$ for \arcA. The spectra are smoothed using a boxcar with a width of 5 pixels with the IRAF task {\tt splot}. In the color coadd image in the right panel, the feature of interest is labeled ``\arcA.'' The scale bar shows the size of the image. The image is oriented north up, east left.}
\label{fig:SystemI}
\end{figure*}
}

\def \tableobservationlog
{
\begin{table*}
  \caption{Spectroscopic Observation Log  \label{table:observationlog}} 
  \centering 
  \footnotesize
  \begin{tabularx}{\linewidth}{*{6}{>{\centering\arraybackslash}X}}
  \hline\hline             
    \DESSystemName \T 	&  UT Date  		& Grating 	& Integration Time	& Seeing 	 \\ 
               			&  					&         	&     (hours)    	&  ($\arcsec$) \\
\toprule
	\SystemA 			& 2015 Oct 11, 12 	& R400/B600 & 0.93/1.00 		& 1.11/1.47 \\
	\SystemB 			& 2015 Oct 11 		& R400/B600 & 1.87/1.00 		& 1.83/1.55 \\
	\SystemC 			& 2015 Oct 11  		& R400 		& 1.00 				& 1.42 \\
	\SystemI 			& 2015 Oct 8, 9 	& R400/B600 & 1.67/1.00 		& 1.83/- \\
	\SystemI 			& 2016 Dec 1, 5 	& R400/B600 & 1.00/1.00 		& 0.73/0.63 \\
	\SystemD 			& 2015 Dec 8 		& R400/B600 & 0.93/1. 			& 0.87/0.81 \\
	\SystemH 			& 2016 Dec 1, 3 	& R400/B600 & 1.00/1.00 		& 0.8/0.85 \\
	\SystemE 			& 2015 Oct 11 		& R400 		& 1.00 				& 1.83 \\
	\SystemF 			& 2015 Oct 12 		& R400		& 1.00 				& 1.3 \\
	\SystemG 			& 2015 Dec 7 		& R400 		& 1.00 				& 0.76 \\
 \bottomrule
\end{tabularx}
\begin{flushleft}
{\bf Notes. } Observation log for follow-up spectroscopic observations. \DESSystemName\ is  derived from the RA and Dec. position of the lensing object at the center of the system.  The seeing is taken from the Gemini Observation Logs, which keep environmental conditions in two-hour increments: the seeing for each exposure set is the average of the seeing recorded at the time steps in the Gemini environmental logs that bracket the observations of interest.
\end{flushleft}
\end{table*}
\normalsize
}

\def \tablelensobjects
{
\begin{table*}
  \caption{Properties of Lensing Systems: Positions and Photometry of Lenses and Source Images \label{table:lensingobjects}} 
  \centering 
  \scriptsize
  \begin{tabularx}{\textwidth}{*{1}{X} *{2}{>{\centering\arraybackslash}X} *{1}{>{\centering\arraybackslash}p{8.9cm}} }
    \hline\hline             
	\ObjectName\  \T  	& R.A. (J2000)	& Dec. (J2000)	& ($g$,$r$,$i$,$z$,$Y$)    	\\ 
						& (deg)      	& (deg)         &     						\\ 
  \toprule
	\SystemA 			& 10.276756 	& -41.926677 	& ($24.55\pm00.28$, $22.39\pm00.04$, $21.12\pm00.02$, $20.62\pm00.02$, $20.22\pm00.10$)\M \\
	A1 					& 10.276189		& -41.924554 	& ($22.65\pm00.05$, $22.16\pm00.03$, $22.15\pm00.05$, $21.98\pm00.07$, $21.82\pm00.42$)\\
	A2 					& 10.274828		& -41.925309 	& ($22.73\pm00.05$, $22.12\pm00.03$, $21.72\pm00.03$, $21.43\pm00.04$, $20.99\pm00.19$)\\
	A3 					& 10.274230		& -41.926661 	& ($22.77\pm00.06$, $22.30\pm00.03$, $21.85\pm00.04$, $21.73\pm00.06$, $21.92\pm00.45$)\\
\hline
	\SystemB 			& 16.098363		& -53.699436 	& ($23.42\pm00.05$, $21.45\pm00.01$, $20.31\pm00.01$, $19.85\pm00.01$, $19.93\pm00.04$)\M \\
	A1* 				& 16.097533		& -53.699283 	& ($24.22\pm00.08$, $22.85\pm00.04$, $21.95\pm00.03$, $21.36\pm00.02$, $21.42\pm00.02$)\\
	A2 					& 16.098237		& -53.700129 	& ($24.14\pm00.10$, $23.08\pm00.05$, $22.46\pm00.05$, $21.86\pm00.06$, $22.22\pm00.28$)\\
\hline
	\SystemC 			& 20.175979		& -51.731407 	& ($22.52\pm00.02$, $20.82\pm00.01$, $19.97\pm00.01$, $19.54\pm00.01$, $19.50\pm00.02$)\M \\
	A1 					& 20.177571		& -51.731638 	& ($21.91\pm00.01$, $21.65\pm00.01$, $21.44\pm00.02$, $21.00\pm00.02$, $21.03\pm00.08$)\\
	A2 					& 20.176720		& -51.730802 	& ($21.83\pm00.01$, $21.43\pm00.01$, $21.16\pm00.01$, $20.69\pm00.02$, $20.73\pm00.06$)\\
	A3 					& 20.175358		& -51.730461 	& ($22.03\pm00.01$, $21.71\pm00.02$, $21.46\pm00.02$, $21.03\pm00.03$, $21.11\pm00.08$)\\
\hline
	\SystemI 			& 36.790976 	& -45.275305 	& ($20.45\pm00.02$, $18.75\pm00.01$, $18.10\pm00.01$, $17.75\pm00.01$, $17.54\pm00.02$)\M \\
	A1 					& 36.792491	 	& -45.274846 	& ($22.58\pm00.05$, $21.79\pm00.04$, $21.22\pm00.04$, $20.93\pm00.06$, $20.65\pm00.13$)\\
\hline
	\SystemD 			& 59.307522		& -47.946327 	& ($20.66\pm00.00$, $19.07\pm00.00$, $18.58\pm00.00$, $18.14\pm00.00$, $18.03\pm00.01$)\M \\
	A1* 				& 59.304217		& -47.946867 	& ($23.14\pm00.05$, $22.24\pm00.03$, $21.69\pm00.02$, $21.35\pm00.02$, $21.27\pm00.02$)\\
	A2* 				& 59.305462		& -47.948650 	& ($22.57\pm00.04$, $21.32\pm00.02$, $20.80\pm00.02$, $20.49\pm00.01$, $20.31\pm00.01$)\\
	A3* 				& 59.307237		& -47.949167 	& ($22.48\pm00.03$, $21.60\pm00.02$, $21.06\pm00.02$, $20.72\pm00.02$, $20.60\pm00.01$)\\
\hline
	\SystemH 			& 64.541167 	& -54.959729 	& ($23.51\pm00.06$, $21.84\pm00.02$, $20.95\pm00.01$, $20.50\pm00.02$, $20.40\pm00.05$)\M \\
	A1 					& 64.540461 	& -54.959360 	& ($22.09\pm00.02$, $21.74\pm00.01$, $21.45\pm00.02$, $21.07\pm00.03$, $20.99\pm00.09$)\\    
\hline
	\SystemE 			& 318.484646 	& -1.240613 	& ($22.51\pm00.02$, $20.97\pm00.01$, $20.28\pm00.01$, $19.89\pm00.01$, $19.99\pm00.04$)\M \\
	A1 					& 318.483975 	& -1.240611 	& ($23.06\pm00.04$, $22.40\pm00.03$, $22.05\pm00.03$, $21.74\pm00.04$, $22.06\pm00.28$)\\
	A2 					& 318.484287	& -1.240847 	& ($21.60\pm00.01$, $21.07\pm00.01$, $20.67\pm00.01$, $20.42\pm00.01$, $20.72\pm00.08$)\\
\hline
	\SystemF 			& 350.367937 	& -46.513625 	& ($23.24\pm00.05$, $21.53\pm00.02$, $20.54\pm00.01$, $20.05\pm00.01$, $20.06\pm00.03$)\M \\
	A1 					& 350.369406 	& -46.514079 	& ($23.26\pm00.05$, $22.45\pm00.04$, $21.75\pm00.03$, $21.26\pm00.03$, $21.26\pm00.11$)\\
	A2 					& 350.368353 	& -46.512958 	& ($22.87\pm00.03$, $22.26\pm00.03$, $21.59\pm00.03$, $21.16\pm00.03$, $21.06\pm00.08$)\\
\hline
	\SystemG 			& 357.375254 	& -51.227520 	& ($22.01\pm00.01$, $19.91\pm00.00$, $19.28\pm00.00$, $18.91\pm00.00$, $18.78\pm00.02$)\M \\
	A1* 				& 357.374425 	& -51.228842 	& ($23.43\pm00.05$, $22.69\pm00.04$, $22.40\pm00.03$, $21.94\pm00.03$, $21.56\pm00.02$)\\
	A2 					& 357.373730 	& -51.227069 	& ($22.45\pm00.02$, $21.82\pm00.02$, $21.53\pm00.02$, $21.15\pm00.02$, $20.86\pm00.12$)\\
\hline
\bottomrule
\end{tabularx} \\
\begin{flushleft}
{\bf Notes. } Positions and photometry of objects for each confirmed strong lensing system. The Object Name refers to each system (\DESSystemName) or lensed source image (e.g., A1) discussed in the text, in Fig.'s~\ref{fig:multipanel} and Fig.'s~\ref{fig:SystemA}-\ref{fig:SystemG}. All positions (R.A., Dec. in J2000) and magnitudes (with \texttt{Mag\_Auto} derived from SourceExtractor) are drawn from the DESDM database, except for some source images that were too faint to detect or were blended with nearby objects in the automated data reduction process---these are marked by an asterisk '*.' For these objects, we measured aperture magnitudes via the Graphical Astronomy and Image Analysis software tool \citep{gaiawebsite}. \nord{Also note that \texttt{MAG AUTO} suffers from segmentation issues and so the magnitudes cannot be used for stellar mass estimates.}
\end{flushleft}
\end{table*}
\normalsize
}

\def \tablerankobserve
{
\begin{table*}
\caption{Observing Strategy \label{table:rankobserve}} 
\centering 
\footnotesize
\begin{tabularx}{\linewidth}{*{3}{>{\centering\arraybackslash}X}}
\hline\hline             
Surface Brightness  		& Grating 		& Integration time 	\\ 
(${\rm mag}/\sqarcsec$)   	& 	      		& (hours) 		  	\\  [0.5ex]  
\toprule
$i_{\rm SB} < 23 $        	& R400 			& 1				  	\\
$23 < i_{\rm SB} < 24$ 		& R400/B600 	& 3.7/1 		  	\\
$i_{\rm SB} >24$			& B600 			& 1	  			    \\  [1ex]
\hline  \hline    
\end{tabularx} 
\begin{flushleft}
{\bf Notes. } Follow-up observing strategy. We report the surface brightness, grating, and integration time for each grouping of follow-up objects.   
\end{flushleft}
\end{table*}
\normalsize
}

\def \tablelensingfeatures
{
\begin{table*}
\footnotesize
\caption{Lensing Features \label{table:lensingfeatures}} 
\begin{center}
\begin{tabularx}{\linewidth}{X *{1}{>{\centering}p{5cm}} *{3}{>{\centering\arraybackslash}X}}
\hline \hline 
\ObjectName\ \T & Spectral Features & Redshift 				& Separation 				& Enclosed Mass   				\\
		  		& 				  	& $\zlens$ or $\zsrc$   & $\thetaesep$ ($\arcsec$) & $\menclosed$ ($\msol$) 		\\    
\toprule
\SystemA 		& ... 				& $\redshiftRMA$ 		& ... 					& ...\M 						\\
A1 				& Ly$\alpha$ 		& $2.5619 \pm 0.0001$ 	& $7.2 \pm 0.5$ 		& $1.7 \pm 0.5 \times 10^{13}$	\\
A2 				& Ly$\alpha$ 		& $2.5618 \pm 0.0001$ 	& ... 					& ...							\\
A3 				& Ly$\alpha$ 		& $2.5616 \pm 0.0002$ 	& ... 					& ...							\\
\hline
\SystemB 		& ... 				& $\redshiftRMB$ 		& ... 					& ...\M 						\\
A1 				& [OII] 			& $1.2318 \pm 0.0001$ 	& $2.2 \pm 0.4$ 		& $2.4 \pm 1.1 \times 10^{12}$	\\
A2 				& [OII] 			& $1.2318 \pm 0.0001$ 	& ... 					& ...							\\
\hline
\SystemC 		& ... 				& $\redshiftRMC$ 		& ... 					& ...\M 						\\
A1 				& [OII] 			& $1.2955 \pm 0.0001$ 	& $3.4 \pm 0.5$ 		& $4.5 \pm 2.8 \times 10^{12}$	\\
A2 				& [OII] 			& $1.2957 \pm 0.0000$ 	& ... 					& ...							\\
A3 				& [OII] 			& $1.2957 \pm 0.0005$ 	& ... 					& ...							\\
\hline
\SystemI 		& ... 				& $\redshiftRMI$ 		& ... 					& ...\M 						\\
A1 				& [OII] 			& $1.2526 \pm 0.0001$ 	& $4.1 \pm 0.3$ 		& $4.3 \pm 1.3 \times 10^{12}$	\\
\hline
\SystemD 		& ... 				& $\redshiftRMD$ 		& ... 					& ...\M 						\\
A1 				& [OII] 			& $0.9156 \pm 0.0001$ 	& $9.4 \pm 0.9$ 		& $1.4 \pm 0.4 \times 10^{13}$	\\
A2 				& [OII] 			& $0.9156 \pm 0.0001$ 	& ... 					& ...							\\
A3 				& [OII] 			& $0.9155 \pm 0.0001$ 	& ... 					& ...							\\
\hline
\SystemH 		& ... 				& $\redshiftRMH$ 		& ... 					& ...\M 						\\
A1 				& [OII] 			& $1.4236 \pm 0.0001$ 	& $2.0 \pm 0.3$ 		& $1.5 \pm 0.7 \times 10^{12}$	\\
\hline
\SystemE 		& ... 				& $\redshiftRME$ 		& ... 					& ...\M 						\\
A1 				& [OII], 
				[NeIII]3868, 
                H$\delta$, 
                H$\gamma$, H$\beta$, 
                [OIII]4959, 
                [OIII]5007 			& $0.7943 \pm 0.0002$ 	& $2.0 \pm 0.5$ 		& $1.3 \pm 1.1 \times 10^{12}$	\\
A2 				& [OII], 
				[NeIII]3868, 
                H$\delta$, 
                H$\gamma$, H$\beta$, 
                [OIII]4959, 
                [OIII]5007 			& $0.7947 \pm 0.0002$ 	& ... 					& ...							\\
\hline
\SystemF 		& ... 				& $\redshiftRMF$ 	& ... 						& ...\M 						\\
A1 				& CII]2326, 
				[NeIV]2424, 
                [NeV]3346, 
                [OII] 				& $1.7469 \pm 0.0014$ 	& $3.3 \pm 0.7$ 		& $3.8 \pm 1.9 \times 10^{12}$	\\
A2 				& CII]2326, 
				[NeIV]2424, 
                [NeV]3346, 
                [OII] 				& $1.7479 \pm 0.0031$ 	& ... 					& ...							\\
\hline
\SystemG 		& ... 				& $\redshiftRMG$ 	& ... 					& ...\M 						\\
A1 				& [OII] 			& $1.3932 \pm 0.0001$ 	& $4.5 \pm 0.7$ 		& $3.8 \pm 1.4 \times 10^{12}$	\\
A2 				& [OII] 			& $1.3938 \pm 0.0001$ 	& ... 					& ...\\
\hline
\bottomrule
\end{tabularx}
\begin{flushleft}
{\bf Notes. } Lensing features of confirmed systems. We show object names for lenses (\DESSystemName) and image labels for sources (e.g., A1), names of spectral features, photometric redshifts of lenses $\zlens$,  spectroscopic redshifts of sources $\zsrc$, an Einstein radius $\thetae$ for each system, and the resulting enclosed masses $\menclosed$. The principal spectral features are all emission lines. Redshifts for the lenses are the photometric redshifts drawn from the DESDM database, and they are the first redshifts listed for each system: the uncertainties have been multiplied by 1.5 times the original estimate, according to the results of \cite{sanchez14} for estimating uncertainties in DES photometric redshift measurement codes. Redshifts of the sources are those measured from spectroscopic follow-up at Gemini South. All the redshifts for a given system are within sufficient agreement that we measure the enclosed masses using only the spectroscopic redshift of the first source and \nord{the \einsteinradiussub} for the system. \nord{Instead of providing Einstein radii (see Poh et al., 2019, in prep.), here we provide rough estimates of the size of lensed features, computed as the weighted average of image-to-lens separation.}
\end{flushleft}
\end{center}
\normalsize
\end{table*}
}

\def \tablelensobjectsnotconfirmed
{
\begin{table*}
  \caption{Inconclusive and Rejected Candidates \label{table:notlensingobjects}} 
  \centering 
  \scriptsize
  \begin{tabularx}{\textwidth}{*{1}{X} *{3}{>{\centering\arraybackslash}X}}
    \hline\hline             
	\ObjectName\  \T  		& R.A. (J2000)	& Dec. (J2000) 	& Confirmation Status	\\ 
							& (deg)      	& (deg)       	& 						\\ 
  \toprule
	DESJ0205-4038 			& 31.272		& -40.642   	& Inconclusive \\
    DESJ0217-5245 			& 34.309 		& -52.758   	& Rejected     \\
	DESJ0300-5001 			& 45.090 		& -50.025 	   	& Inconclusive \\
    DESJ0322-5234 			& 50.567		& -52.576		& Inconclusive \\
    DESJ0342-5355 			& 55.520		& -53.921   	& Inconclusive \\
    DESJ0428-4409 			& 67.081		& -44.166  		& Rejected     \\
	DESJ0450-5715 			& 72.537		& -57.256		& Inconclusive \\
    DESJ0510-5637 			& 77.555		& -56.632		& Inconclusive \\
    DESJ0536-5338 			& 84.022		& -53.647		& Inconclusive \\
    DESJ0538-4735 			& 84.518		& -47.588		& Inconclusive \\
    DESJ0602-4524 			& 90.693		& -45.412		& Inconclusive \\
    DESJ2127-5149 			& 321.779 		& -51.831 	   	& Inconclusive \\  
\hline
\bottomrule
\end{tabularx} \\
\begin{flushleft}
{\bf Notes. } A list of each system that could not be confirmed as a strong lensing system --- either the data is "inconclusive" for a confirmation or the data indicates the system can be "rejected." The Object Name refers to each system (\DESSystemName). All positions (R.A., Dec. in J2000). DESJ0217-5245 was rejected: the source object appears as a multiply-imaged red galaxy, but these turned out to be a background group. DESJ0428-4409 was also rejected: the blue source galaxy is actually a foreground star-forming galaxy.
\end{flushleft}
\end{table*}
\normalsize
}

\def \figuremultipanelversion {009} 
\def \figuresystemversion {12} 		

\title{Observation and Confirmation of Nine Strong Lensing Systems in Dark Energy Survey Year 1 Data} 

\author[DES Collaboration]{
\parbox{\textwidth}{
\Large
B.~Nord,$^{1, 2, 3}$
E.~Buckley-Geer,$^{1}$
H.~Lin,$^{1}$
N.~Kuropatkin,$^{1}$
T.~Collett,$^{4}$
D.~L.~Tucker,$^{1}$
H.~T.~Diehl,$^{1}$
A.~Agnello,$^{5}$
A.~Amara,$^{6}$
T.~M.~C.~Abbott,$^{7}$ 
S.~Allam,$^{1}$
J.~Annis,$^{1}$
S.~Avila,$^{4}$
K.~Bechtol,$^{8,9}$
D.~Brooks,$^{10}$
D.~L.~Burke,$^{11,12}$
A.~Carnero~Rosell,$^{13,14}$
M.~Carrasco~Kind,$^{15,16}$
J.~Carretero,$^{17}$
C.~E.~Cunha,$^{11}$
L.~N.~da Costa,$^{14,18}$
C.~Davis,$^{11}$
J.~De~Vicente,$^{13}$
P.~Doel,$^{10}$
T.~F.~Eifler,$^{19,20}$
A.~E.~Evrard,$^{21,22}$
E.~Fernandez,$^{17}$
B.~Flaugher,$^{1}$
P.~Fosalba,$^{23,24}$
J.~Frieman,$^{1,2}$
J.~Garc\'ia-Bellido,$^{25}$
E.~Gaztanaga,$^{23,24}$
D.~Gruen,$^{26,11,12}$
R.~A.~Gruendl,$^{15,16}$
G.~Gutierrez,$^{1}$
W.~G.~Hartley,$^{10,6}$
D.~L.~Hollowood,$^{27}$
K.~Honscheid,$^{28,29}$
B.~Hoyle,$^{30,31}$
D.~J.~James,$^{32}$
K.~Kuehn,$^{33}$
N.~Kuropatkin,$^{1}$
O.~Lahav,$^{10}$
M.~Lima,$^{34,14}$
M.~A.~G.~Maia,$^{14,18}$
M.~March,$^{35}$
J.~L.~Marshall,$^{36}$
P.~Melchior,$^{37}$
F.~Menanteau,$^{15,16}$
R.~Miquel,$^{38,17}$
A.~A.~Plazas,$^{20}$
A.~K.~Romer,$^{39}$
A.~Roodman,$^{11,12}$
E.~S.~Rykoff,$^{11,12}$
E.~Sanchez,$^{13}$
V.~Scarpine,$^{1}$
R.~Schindler,$^{12}$
M.~Schubnell,$^{22}$
I.~Sevilla-Noarbe,$^{13}$
M.~Smith,$^{40}$
M.~Soares-Santos,$^{41}$
F.~Sobreira,$^{42,14}$
E.~Suchyta,$^{43}$
M.~E.~C.~Swanson,$^{16}$
G.~Tarle,$^{22}$
D.~Thomas,$^{4}$
and Y.~Zhang$^{1}$
\begin{center} (DES Collaboration) \end{center}
}
\vspace{0.4cm}
\\
\parbox{\textwidth}{
$^{1}$ Fermi National Accelerator Laboratory, P. O. Box 500, Batavia, IL 60510, USA\\
$^{2}$ Kavli Institute for Cosmological Physics, University of Chicago, Chicago, IL 60637, USA\\
$^{3}$ Department of Astronomy and Astrophysics, University of Chicago, Chicago, IL 60637, USA\\
$^{4}$ Institute of Cosmology and Gravitation, University of Portsmouth, Portsmouth, PO1 3FX, UK\\
$^{5}$ European Southern Observatory, Karl-Schwarzschild-Strasse 2, 85748 Garching, Germany \\
$^{6}$ Department of Physics, ETH Zurich, Wolfgang-Pauli-Strasse 16, CH-8093 Zurich, Switzerland\\
$^{7}$ Cerro Tololo Inter-American Observatory, National Optical Astronomy Observatory, Casilla 603, La Serena, Chile\\
$^{8}$ LSST, 933 North Cherry Avenue, Tucson, AZ 85721, USA\\
$^{9}$ Physics Department, 2320 Chamberlin Hall, University of Wisconsin-Madison, 1150 University Avenue Madison, WI  53706-1390\\
$^{10}$ Department of Physics \& Astronomy, University College London, Gower Street, London, WC1E 6BT, UK\\
$^{11}$ Kavli Institute for Particle Astrophysics \& Cosmology, P. O. Box 2450, Stanford University, Stanford, CA 94305, USA\\
$^{12}$ SLAC National Accelerator Laboratory, Menlo Park, CA 94025, USA\\
$^{13}$ Centro de Investigaciones Energ\'eticas, Medioambientales y Tecnol\'ogicas (CIEMAT), Madrid, Spain\\
$^{14}$ Laborat\'orio Interinstitucional de e-Astronomia - LIneA, Rua Gal. Jos\'e Cristino 77, Rio de Janeiro, RJ - 20921-400, Brazil\\
$^{15}$ Department of Astronomy, University of Illinois at Urbana-Champaign, 1002 W. Green Street, Urbana, IL 61801, USA\\
$^{16}$ National Center for Supercomputing Applications, 1205 West Clark St., Urbana, IL 61801, USA\\
$^{17}$ Institut de F\'{\i}sica d'Altes Energies (IFAE), The Barcelona Institute of Science and Technology, Campus UAB, 08193 Bellaterra (Barcelona) Spain\\
$^{18}$ Observat\'orio Nacional, Rua Gal. Jos\'e Cristino 77, Rio de Janeiro, RJ - 20921-400, Brazil\\
$^{19}$ Department of Astronomy/Steward Observatory, 933 North Cherry Avenue, Tucson, AZ 85721-0065, USA\\
$^{20}$ Jet Propulsion Laboratory, California Institute of Technology, 4800 Oak Grove Dr., Pasadena, CA 91109, USA\\
$^{21}$ Department of Astronomy, University of Michigan, Ann Arbor, MI 48109, USA\\
$^{22}$ Department of Physics, University of Michigan, Ann Arbor, MI 48109, USA\\
$^{23}$ Institut d'Estudis Espacials de Catalunya (IEEC), 08034 Barcelona, Spain\\
$^{24}$ Institute of Space Sciences (ICE, CSIC),  Campus UAB, Carrer de Can Magrans, s/n,  08193 Barcelona, Spain\\
$^{25}$ Instituto de Fisica Teorica UAM/CSIC, Universidad Autonoma de Madrid, 28049 Madrid, Spain\\
$^{26}$ Department of Physics, Stanford University, 382 Via Pueblo Mall, Stanford, CA 94305, USA\\
$^{27}$ Santa Cruz Institute for Particle Physics, Santa Cruz, CA 95064, USA\\
$^{28}$ Center for Cosmology and Astro-Particle Physics, The Ohio State University, Columbus, OH 43210, USA\\
$^{29}$ Department of Physics, The Ohio State University, Columbus, OH 43210, USA\\
$^{30}$ Max Planck Institute for Extraterrestrial Physics, Giessenbachstrasse, 85748 Garching, Germany\\
$^{31}$ Universit\"ats-Sternwarte, Fakult\"at f\"ur Physik, Ludwig-Maximilians Universit\"at M\"unchen, Scheinerstr. 1, 81679 M\"unchen, Germany\\
$^{32}$ Harvard-Smithsonian Center for Astrophysics, Cambridge, MA 02138, USA\\
$^{33}$ Australian Astronomical Optics, Macquarie University, North Ryde, NSW 2113, Australia\\
$^{34}$ Departamento de F\'isica Matem\'atica, Instituto de F\'isica, Universidade de S\~ao Paulo, CP 66318, S\~ao Paulo, SP, 05314-970, Brazil\\
$^{35}$ Department of Physics and Astronomy, University of Pennsylvania, Philadelphia, PA 19104, USA\\
$^{36}$ George P. and Cynthia Woods Mitchell Institute for Fundamental Physics and Astronomy, and Department of Physics and Astronomy, Texas A\&M University, College Station, TX 77843,  USA\\
$^{37}$ Department of Astrophysical Sciences, Princeton University, Peyton Hall, Princeton, NJ 08544, USA\\
$^{38}$ Instituci\'o Catalana de Recerca i Estudis Avan\c{c}ats, E-08010 Barcelona, Spain\\
$^{39}$ Department of Physics and Astronomy, Pevensey Building, University of Sussex, Brighton, BN1 9QH, UK\\
$^{40}$ School of Physics and Astronomy, University of Southampton,  Southampton, SO17 1BJ, UK\\
$^{41}$ Brandeis University, Physics Department, 415 South Street, Waltham MA 02453\\
$^{42}$ Instituto de F\'isica Gleb Wataghin, Universidade Estadual de Campinas, 13083-859, Campinas, SP, Brazil\\
$^{43}$ Computer Science and Mathematics Division, Oak Ridge National Laboratory, Oak Ridge, TN 37831\\
}
}

\begin{document}

\maketitle 


\begin{abstract}
We describe the observation and confirmation of \nbconfirmtext\ new strong gravitational lenses discovered in Year 1 data from the Dark Energy Survey (DES). 
We created candidate lists based on a) galaxy group and cluster samples and b) photometrically selected galaxy samples. 
We  selected \nbcandidate\ candidates through visual inspection and then used the Gemini Multi-Object Spectrograph (GMOS) at the Gemini South telescope to acquire spectroscopic follow-up of \nbfollowup\ of these candidates. 
Through analysis of this spectroscopic follow-up data, we confirmed \nbconfirmtext\ new lensing systems and rejected \nbrejecttext\ candidates, but the analysis was inconclusive on \nbuncertain\ candidates. 
For each of the confirmed systems, we report measured spectroscopic properties, estimated \einsteinradiussub, and estimated enclosed masses. 
The sources that we targeted have an {\it i}-band surface brightness range of $i_{\rm SB} \sim 22 - 24\, {\rm mag}/\sqarcsec$ and a spectroscopic redshift range of  $\zspec \sim 0.8 - 2.6$. The lens galaxies have a photometric redshift range of $\zlens \sim 0.3 - 0.7$. 
The lensing systems range in \einsteinradiussub\ $2 - 9\arcsec$ and in enclosed mass  $10^{12} - 10^{13} \msol$.
\end{abstract}


\section{introduction}
Strong gravitational lensing uniquely demonstrates the interplay between energy and space-time. 
The light rays from background sources, like galaxies and quasars, are deflected by massive dark matter halos that are in alignment along an observer's line of sight, producing highly distorted images of these sources.  
In this role, strong lensing provides opportunities to study multiple astrophysical and cosmological phenomena --- from the evolution of highly magnified distant galaxies in the early universe to the matter distribution on the scales of galaxies, as well as cosmic acceleration \citep[e.g.,][]{treu10}.

About a thousand lensing systems have been discovered over the last few decades, which we detailed in \cite{nord16}. Modern large-scale surveys have the potential to increase this sample by orders of magnitude by virtue of their relatively large depth and sky area.
For example, based on selection function estimates of spectroscopic confirmations \citep{nord16} and simulation-based forecasts of future surveys \citep{collett12}, \nord{the Dark Energy Survey \citep[DES\footnote{\url{darkenergysurvey.org}};][]{diehl14,flaugher15,2016MNRAS.460.1270D}} is likely to observe more than 2000 galaxy- and cluster-scale lensing systems with arcs, and over a hundred lensed quasars \citep{oguri10}. 
The Large Synoptic Survey Telescope \citep[LSST\footnote{\url{lsst.org}};][]{iveziclsst08} data will contain an order of magnitude more \citep{oguri10, collett15}. 

Each species of strong lens --- characterized by the scale of the lens, the objects being lensed, and image morphology --- has particular strengths for studying dark matter, dark energy, the cosmic expansion rate, galaxy evolution, and other phenomena.
For example, modeling individual lens systems can constrain mass profiles and mass-to-light ratios of early-type galaxies \citep[e.g.,][]{sonnenfeld2013}. 
Modeling the population of galaxy- and group-scale lenses can provide constraints on profiles of dark matter haloes in lenses \citep[e.g.,][]{more12,more16}. 
Double-source-plane lenses --- when there are two sources along the line of sight behind a lens --- can constrain cosmic dark matter and dark energy densities, largely independent of the Hubble constant \citep{collett15, linder16}; to date only a few have been discovered \citep[e.g.,][]{gavazzi08, tanaka16}. 
Time-delay cosmology uses variable objects, like lensed quasars and supernovae to measure the Hubble rate, in a manner largely independent of dark matter and dark energy densities \citep{refsdal64, blandford92} or the dark energy equation of state \citep{linder11}. 
Recent work by the H0liCoW ($H_0$ Lenses in COSMOGRAIL's Wellspring) collaboration provide constraints on the Hubble constant with precision that is competitive with other standard cosmological probes \citep{suyu17, bonvin16}.

Lenses of different kinds have been the object of targeted searches in DES, including quasar lenses \citep{agnello15, agnello17, lin17, ostrovski17}, as well as galaxy-galaxy lenses \citep{nord16}. 
The lensed quasar search is the focus of the STRong-lensing Insights into Dark Energy Survey (STRIDES) program\footnote{\url{http://strides.astro.ucla.edu/}}.
The current work extends this line of DES-wide investigation, with galaxy-galaxy lenses spanning different environments: isolated galaxies, groups, and clusters.
\nord{Spectroscopic confirmation is a time-consuming, yet critical process for astrophysical and cosmological analyses.
While thousands of candidates may be observable with DES, the sample size that is optimized for the aforementioned science goals is constrained by observational resources and the number of systems with science potential.}

In this work, we describe the spectroscopic confirmation of \nbconfirmtext\ new strong lensing systems discovered in DES Year 1 (Y1) data. 
We selected these systems based on the brightness of source galaxies, as well as on the potential for modeling the system --- i.e., morphological simplicity and potential to examine the mass profile of the underlying dark matter halo.
We first created lists of candidates based on a) samples of galaxy groups and clusters and b) photometrically selected galaxies --- totaling $\sim9500$ candidates.
We then visually inspected an image of each candidate system. 
This led to the selection of \nbcandidate\ candidates, \nord{and we performed spectroscopic follow-up on \nbfollowup\ of those. We then analyzed the spectroscopic data to determine the redshifts of the sources in those lenses to discern if the systems are indeed strong lenses. We describe in more detail the lens candidate search process in \S\ref{sec:search}.}

In this work, we focus on \nbconfirmtext\ systems (shown in Fig.~\ref{fig:multipanel}) that show evidence --- morphology, photometry, and spectroscopy --- of strong gravitational lensing: three of the systems are galaxy-scale lenses, five are group-scale and one is a cluster-scale lens. 
All of them are newly discovered and confirmed objects. 
Detailed mass models of a subset of the confirmed systems will be presented in a separate paper \citep{poh17}.

The paper has the following structure.
We describe the DES Y1 data in \S\ref{sec:data}.
We discuss the search for lens candidates in \S\ref{sec:search} and then the follow-up spectroscopic observations and analysis in \S\ref{sec:follow-up}. 
We present the \nbconfirmtext\ confirmed systems and their properties in \S\ref{sec:sample}. 
We then conclude in \S\ref{sec:summary}. 
The AB system is used for all magnitudes. 
A Planck $\Lambda$CDM cosmology with spatially-flat priors is assumed: $\Omega_{\rm M}=0.308$, $\Omega_{\Lambda}=0.692$, and $H_0=67.8\, \kms\,{\rm Mpc}^{-1}$ \citep{planckcosmo15}.

\section{DES Year 1 Data}\label{sec:data}

DES covers $\sim5000$ \sqdeg\ in a deep, wide-field survey of the southern Galactic Cap in five optical filters ($grizY$). 
It extends to a depth of $i\sim24$ mag at $10\sigma$ detection threshold and a 27\sqdegadjective\ supernova survey across 10 fields --- two deep and eight shallow. 
The depth in the repeated supernova area is typically two magnitudes deeper than the wide-field survey. 
The survey is undertaken with the Dark Energy Camera \citep[DECam][]{flaugher15}, which is a 3-sq.-deg. CCD mosaic camera mounted on the Blanco 4m telescope at the Cerro Tololo Inter-American Observatory (CTIO) in the Chilean Andes. 
The DES footprint is observable between August and mid-February. 
Each year, DES is allocated $\sim105$ nights, and it has now completed all five and a half years of planned observations. 
Because of its large field of view and red-sensitive CCDs, DECam is particularly suited to high-redshift survey work

For this work, we use the Year One First Annual (Y1A1) internal collaboration release of the DES data. 
The survey and operations are described in \citet{diehl14}. 
The data in the Y1A1 release were acquired between 2013 August 15th and 2014 February 9th. 
\nord{These data cover $\sim1840$ \sqdeg\ to a median $10\sigma$ point source depth calculated with Source Extractor's fixed-aperture magnitudes (MAG APER), with 1.95`` apertures, of 24.19, 23.85, 23.25, 22.55, and 21.20 in the $g$, $r$, $i$, $z$, and $Y$ bands, respectively \citep{2018ApJS..235...33D}.}

\nord{The reduction of the images from DES Y1 data was performed by the DES data management (DESDM) team \citep{2018PASP..130g4501M}.}  
After detrending, the single-epoch images were combined into `coadd tiles' after first being calibrated and background-subtracted.
The tiles are coadd images comprising one to five exposures in each of the five wavelength bandpasses.
On average, in each tile, the coverage comprised 3.5 exposures. 
Each coadd tile has dimensions $0.73\ {\rm deg.}\times 0.73\ {\rm deg.}$, which are defined to cover the full footprint of DES homogeneously. 
The final survey depth is deeper than the Y1A1 release, which consists of 3707 coadded tiles.
This footprint covers two non-contiguous regions: one overlaps the deeply imaged Stripe 82 \citep{jiang14} from SDSS \citep{york00, abazajian09}; the other overlaps the area that is covered by the South Pole Telescope \citep[SPT;][]{carlstrom11}. 
More details of the reductions are available in \citet{drlicawagner17arxiv}. 

Source Extractor \citep[][]{bertin96, bertin11} image detection software is used for catalog source detection. 
It is deployed in double-image mode, and for the detection image, uses the $\chi^2$ detection image, which is constructed from the combination of the $r$, $i$, and $z$ band images. 
Positional and photometric data of all the objects in this study (see Table~\ref{table:lensingobjects}) come from this object catalog. 
While MAG APER is used for the measures of depth in the Y1A1 catalog release, in this work, we use the magnitude measure, MAG AUTO to perform catalog searches for lens candidates. 
 
\section{Lens Search}
\label{sec:search}

\nord{To identify candidates in DES Y1 data, we used a) samples of galaxy clusters selected via DES data or via SPT data, and b) galaxies selected based on photometry.} 
A number of expert scanners visually inspected candidates in search of visually compelling evidence (e.g., morphology, color, brightness) that the systems would be useful for mass modeling. 


\nord{The search method used to create candidate lists for visual inspection selects primarily for blue or red source galaxies near single galaxies or within groups or clusters of galaxies.}
Searches of 7328 optically selected galaxy clusters --- found with the \redmapper\ algorithm \citep{rykoff14} ---  as well as a selection of galaxies from the \redmagic\ catalog \citep{rozo2016,cawthon2017} yielded 374 candidates in the DES Y1 footprint. This is described in \S~3.3 of \citet{diehl17pub}, which will from here on be referred to as \diehlpaper. 
A third search of SPT galaxy clusters yielded 66 more candidates (described in \S3.1 \diehlpaper). 
Candidates from searches described elsewhere in \diehlpaper\ added 88 more. 
Some candidates were the result of serendipitous discovery, and those were not described in \diehlpaper. 

The combined searches resulted in a list of 112 candidate systems. 
A sample of 46 candidates were selected on the bases of 1) science cases (e.g., the potential for modeling the dark matter halo of the lens itself) and 2) objects that are already targeted by other follow-up efforts (e.g., some SPT clusters).
Note that the ranking discussed in \diehlpaper\ is based purely on how likely an object is to be a lens (not on how easy it would be to follow up).

We cross-checked both our follow-up candidate sample and the full Y1A1 footprint against the Master Lens Database (MLDB; last updated 2 February 2018), which contains 674 candidates and confirmed systems. None of the candidates we observed during our follow-up for this paper are found in the MLDB. 

There are two areas in the DES Y1A1 footprint: one overlaps with SPT, and the other overlaps with SDSS Stripe 82. From MLDB, 202 lenses are within the Y1A1 footprint --- 28 in the SPT area and 174 in Stripe82 area. Those in the SPT area are sufficiently high redshift to not be visible within wavelengths observed by DES. Those in the Stripe82 area were already discovered and are not within our follow-up sample. 

\section{Spectroscopic Follow-up at Gemini/GMOS}
\label{sec:follow-up}

In this paper, we aim to report the spectroscopic evidence for strong gravitational lensing in a sample of candidates that were selected through visual scans of images.
We identified \nbcandidate\ candidates during the lens search, from which we chose the \nbfollowup\ candidates that are most suitable for follow-up with spectroscopic observations and analysis.
The bases for this down-selection are 1) brightness of source galaxies and 2) suitability of system for observation (e.g., mask alignment). 
\nord{We measure the spectroscopic redshifts of source galaxies to determine if they are larger than the putative lens redshifts.}

We require a wide spectral range to search for patterns of narrow emission lines. 
We expect some sources to be late-type emission-line galaxies.
With these, we look for several features, such as \OIII\ and \Hbeta\ to $z\sim 1.0$; \Hdelta, \Hgamma, and \OIIdoublet\footnote{Here, we refer to the doublet, because our resolution is not sufficient to resolve both lines in the doublet.} to $z\sim 1.7$; and \Lya\ in the range $z\sim 2.7 - 7.2$.

As part of the Gemini Large and Long Program (GS-2015B-LP-5 and GS-2016B-LP-5)\footnote{\url{http://www.gemini.edu/?q=node/12238\#Buckley}}, we used the multi-object mode of the Gemini Multi-Object Spectrograph \citep[GMOS;][]{2004PASP..116..425H} the Gemini South Telescope to spectroscopic observations of our candidates.
\nord{This proposal includes the goal of creating a spectroscopic sample of red galaxies for photometric redshift calibration.} 
Below, we describe the observing strategy and the reductions of the spectroscopic data.

\subsection{Observing Strategy}
\label{sec:followup:observingstrategy}

We used the following procedure for planning the follow-up. 
For the purposes of planning the follow-up, the  were 
We first ranked \nbfollowup\ candidates by their $i$-band surface brightness, which is calculated using an aperture that is $2\arcsec$ in diameter.
We defined three sets of gratings and exposure times. 
The combinations depend on these surface brightness classes:
for objects with surface brightness $i_{\rm SB} < 23 $, we integrate for 1 hour in the R400 grating; 
for objects with surface brightness $23 < i_{\rm SB} < 24$, we integrate for 3.7 hours in the R400 grating and then for 1 hour in the B600 grating;
for objects with surface brightness, $i_{\rm SB} >24$, we integrate for 1 hour in the B600 grating.
See Table~\ref{table:observationlog} for details of the observations that were performed.

We mostly performed 1-hour observations, which allowed us to obtain sufficient signal in the cases for which a lens could be confirmed in that amount of time. 
More than one hour of integration would be unlikely to yield enough additional signal-to-noise to warrant spending the time.
In the event that a clear signal did not appear during the prescribed observations, we chose to not perform additional observations or longer integration times for any system. 
We made this decision to conserve telescope time and to maximize the number of strong lens system confirmations.
Moreover, consistent integration time across the observation fields incur consistent depth---a requirement for the photo-z calibration targets.

We centered each field's mask on the lens of the candidate system, and we placed slists on the images of the sources.
In some cases, we shifted the center of the field to accommodate a suitable guide star.
We also rotated the slit mask (i.e., rotated teh position angle of the system) to include as many source targets as possible. 
Slits were first placed on as many of the source images as possible. 
Then, for the calibration of DES photometric redshifts, the unused slits (about 40) were placed on galaxy targets. 
While the slit lengths varied, all the slits were $1\arcsec$ in width. 
This setting accommodated both the object and the amount of sky sufficient for reliable background subtraction.
In some cases, the slits were tilted to maximize the flux captured from an extended source. 

To obtain spectra of objects with redshift $z<1.7$ and wavelength coverage $\sim5000-10000$\AA, we use the R400 grating in conjunction with the GG455 filter. 
For spectra of objects with redshift $z>2$ and within the wavelength range $3250-6250$\AA, we use the B600 grating without a filter.

For an observing sequence with a 1-hour science integration, we first took a pair of half-hour 900-second science exposures.
Then, we used a Quartz-Halogen lamp to take a flat field and a Quartz-Halogen lamp to take a calibration spectrum. 
To cover the gap between the CCDs, we then shifted to a different central wavelength. 
Finally, we repeated this sequence in reverse order. 
For the The 3.7-hour science integration, we instead used 840-second science exposures and repeated the above sequence 16 times.
We facilitate the removal of cosmic rays by dividing the integration time into multiple exposures. 
We binned the data $2\times2$, which gave effective dispersions of $1.0$ and $1.5$ \AA/pixel for the B600 and R400 gratings, respectively. 

\subsection{Spectroscopic Reductions}
\label{sec:followup:reductions}

We used the Gemini IRAF package v1.13.1\footnote{\url{http://www.gemini.edu/sciops/data-and-results/processing-software}} for IRAF v2.16 to reduce the exposures. 
Some of the Gemini IRAF tasks were modified to provide additional flexibility in the data reduction. 
First, for each wavelength dither in a given system, we use the {\tt gsflat} task to process (including bias subtraction) the flat field.
We then use these processed flat images and {\tt gsreduce} to reduce each science exposure.
Then, the two exposures are combined with {\tt gemcombine}. 
We then use the {\tt gswavelength} and {\tt gstransform} tasks to perform wavelength calibration and transformation on each dither.

Then we coadd a pair of dithers on the new common wavelength scale, which eliminates CCD chip gaps.
We use {\tt gsextract} (which calls the {\tt apall} task) to perform sky subtraction and 1D spectral extraction.
We use night sky lines from the science spectra to add calibration lines to the 5500\AA-6400\AA\ wavelength range. 
We modified the canonical reduction process by taking the log of the fluxes in the calibration spectra to enable simultaneous automated identification of both the strong lines above 7000\AA\ and the weaker lines below. 
For improved interactive flexibility in this part of the reduction, {\tt gswavelength} and {\tt gsextract} were modified to allow wavelength calibration and 1D spectral extraction, respectively, for selected individual slits, as needed. 
Finally, we use the {\tt emsao} task within the {\tt rvsao} IRAF package \citep{kurtz98} for feature identification and redshift estimation.

\section{Sample of Confirmed Lenses}
\label{sec:sample}

We confirmed that \nbconfirmtext\ of the \nbfollowup\ observed candidate systems are indeed strong lensing systems, and we rejected \nbrejecttext\ candidates. The remaining \nbuncertaintext\ were not confirmed, and spectroscopic analysis was inconclusive for those candidates.

One of the two rejected systems contains a foreground star-forming galaxy, and the other contains a background group (rather than the apparent multiply lensed red galaxy). 
The systems that we failed to confirm exhibit promising lensing features, but the sources have no discernible continuum emission, no spectral features, or both.
The measurements of spectroscopic features are too low signal-to-noise with the integration times in our observing program, or they have redshifts that are outside the range of the optical observations in our observing program (i.e., in the redshift desert). 
We list the rejected and inconclusive systems in Table~\ref{table:notlensingobjects}.

In Fig.~\ref{fig:multipanel}, we show a multi-panel figure of the confirmed systems.
In Table~\ref{table:lensingobjects}, for each observed lensing system, we list the positions and photometry of the candidate lens and source(s). 
The sample is comprised of three galaxy-scale lenses (b, f, g), five group-scale lenses (a, c, d, h, i) and one cluster-scale lens (e).

In this section, we provide details for each system --- the important spectral features, the measured redshifts, and simple mass estimates. 
Fig.'s~\ref{fig:SystemA}-\ref{fig:SystemG} show the reduced 1D spectra and a cut-out of the field centered on the central lensing object, including labeled source positions for each system. 
 
We estimate the enclosed mass $\menclosed$ of the lensing system under the assumption of a singular isothermal sphere (SIS) mass profile \citep{narayan96}:
\begin{equation}
\menclosed = \frac{c^2}{4G} \thetae^2 \left(\frac{D_{\rm L} D_{\rm S}}{D_{\rm L S}}\right), \label{eqn:menclosed}
\end{equation}
where $c$ is the speed of light, $G$ is Newton's gravitational constant, $\thetae$ is the Einstein radius, and $D$ is the angular diameter distance.
In particular, $\DL$, $\DS$, and $\DLS$ are the  angular diameter distances to the lens, to the source, and between the lens and the source, respectively.
\nord{We use the \einsteinradiussub\ (calculated in \diehlpaper) as an approximation for the Einstein radius  $\thetaesep$. 
It is measured by taking the average of the distances between a targeted source image and the selected central lensing galaxy. 
This estimate of the Einstein radius is only accurate to a factor of 2, and it's an overestimate for cases in which only the brightest source image is used to approximate the radius.}

We use the standard deviation of the distances summed in quadrature with the pixel scale of DECam (0.263\arcsec/pix) to estimate the uncertainty in this distance. 
The angular diameter distances only depend on redshifts of the objects and on cosmological parameters. 
Table~\ref{table:lensingfeatures} summarizes key information for all the confirmed systems --- spectral features, photometric redshifts of lenses, spectroscopic redshifts of sources, \einsteinradiussub, and enclosed masses. 
This strategy is similar to that used in the Sloan Bright Arc Survey \citep[SBAS; e.g.,][]{diehl09}.

We combine uncertainties from the measurements of the distances and \einsteinradiussub\ to estimate frequentist uncertainties for the enclosed mass. 
According to the prescription of \citet{sanchez14} for photometric redshifts in DES, the photometric redshift uncertainties have been multiplied by a factor of $1.5$. 
The spectroscopic redshift uncertainties are the result of a sum in quadrature of a) the uncertainty in wavelength calibration and b) the uncertainty in the redshift determination from the IRAF function {\tt emsao}. 
These spectroscopic redshift uncertainties are then propagated to the angular diameter distances. 

The mass uncertainty results from the sum in quadrature of uncertainties from the \einsteinradiussub, and the angular diameter distances. 
The uncertainty on each angular diameter distance scales with the redshift error, which ranges $\sim0.008 - 0.19\%$ for spectroscopic redshifts and $\sim 3.2 - 17.5\%$ for lens redshifts.
The uncertainty in the \einsteinradiussub, which ranges $\sim 6-22\%$ is purely statistical.
This results in mass uncertainties in the range $\sim25-80\%$. 
Table~\ref{table:lensingfeatures} summarizes the lensing features for the confirmed systems. 


A subset of the systems have sources with multiple, detailed images, making them amenable to detailed modeling, which will be performed in a separate paper \citep{poh17}.

\subsection{\SystemA}
\label{sec:sample:systemA}

\SystemA\ is a group-scale system.
The largest central red galaxy has a DESDM photometric redshift of $\zlens=\redshiftlensA$.  
As shown in Fig.~\ref{fig:multipanel}a, there are three prominent blue-pink arcs, \arcA, \arcB, and \arcC\ to the north, northwest, and west, respectively, of the lensing galaxy. 
The right panel of Fig.~\ref{fig:SystemA} also displays these arcs. 
We identify emission lines in the follow-up B600 spectroscopy of the three images near the observed wavelength of $\sim4329$\AA\ in all three spectra (Fig.~\ref{fig:SystemA}, left panel). 
When we account for the absence of spectral features in the R400 spectra, as well as the photometric redshift of the lens galaxy, we can assign spectral features to be \Lya.
This then gives redshifts $\zsrc=\redshiftsrcAA$, $\redshiftsrcAB$, and $\redshiftsrcAC$ for \arcA, \arcB, and \arcC, respectively.

We identified no counter-images.
In the R400 spectrum for \arcA, there is an \OIIdoublet\ line at $\sim6620$\AA (not shown in Fig.~\ref{fig:SystemA}), which corresponds to a redshift of $z=0.7761\pm0.00007$ of a foreground object. This is clearly visible as a bluer galaxy superimposed on \arcA.

We use the \redmapper\ redshift $\zlens=\redshiftRMA$ and the estimated \einsteinradiussub\ of $\thetaesep=\SystemARadius\arcsec$ to calculate an enclosed mass of $\menclosed=\SystemAMass$ for this system.


\subsection{\SystemB}
\label{sec:sample:systemB}

\SystemB\ is a galaxy-scale lensing system. 
The red lensing galaxy has a photometric redshift $\zlens=\redshiftlensB$, and there are two relatively red arcs to the west and south, labeled \arcA\ and \arcB, respectively, as shown in Fig.~\ref{fig:multipanel}b and in Fig.~\ref{fig:SystemB}. 
We identify strong emission lines for \arcA\ and \arcB\ (Fig.~\ref{fig:SystemB}, left panel) in both R400 spectra near an observed wavelength of $\sim8319$\AA. 
In the absence of features in the B600 spectra, we associate this feature with \OIIdoublet, yielding redshifts $\zsrc=\redshiftsrcBA$ and $\redshiftsrcBB$ for \arcA\ and \arcB, respectively. We see no counter-images for this source.

We use the \redmapper\ redshift $\zlens=\redshiftRMB$ and the estimated \einsteinradiussub\ of $\thetaesep=\SystemBRadius\arcsec$ to calculate an enclosed mass of $\menclosed=\SystemBMass$ for this system. 


\subsection{\SystemC}
\label{sec:sample:systemC}

The central red galaxy of this group-scale system \SystemC\ has a photometric redshift $\zlens=\redshiftlensC$. 
There exist three source images, \arcA, \arcB, and \arcC\ to the east, northeast, and north-northwest, respectively, of the central red galaxy in the image, as shown in Fig.~\ref{fig:multipanel}c and in the right panel of Fig.~\ref{fig:SystemC}. 
In all R400 data (Fig.~\ref{fig:SystemC}, left panel), there are emission-line features near $\sim8557$\AA, which we attribute to \OIIdoublet. 
The resulting redshifts for the source images are $\zsrc=\redshiftsrcCA$, $\redshiftsrcCB$, and $\redshiftsrcCC$ for arcs \arcA, \arcB, and \arcC, respectively. 

This lensing system resides in a group environment. 
There is a counter-image at the southwest that lies nearly on top of the central red lensing galaxy. 
We were not able to target it for follow-up spectroscopic observation.

We use the \redmapper\ redshift $\zlens=\redshiftRMC$ and the estimated \einsteinradiussub\ of $\thetaesep=\SystemCRadius\arcsec$ to calculate an enclosed mass of $\menclosed=\SystemCMass$ for this system.


\subsection{\SystemI}
\label{sec:sample:systemI}

\SystemI\ is a group-scale lens with DES photometric redshift, $\zlens=\redshiftlensI$, and one blue knot (\arcA) to the south. 
The system is shown in Fig.~\ref{fig:multipanel}i and in the right panel of Fig.~\ref{fig:SystemI}. 
Near wavelength $\sim8396$\AA, we identify an \OIIdoublet\ emission line in the R400 spectra of the blue knot. 
This yields source redshifts of $\zsrc=\redshiftsrcIA$ \arcA\ (Fig.~\ref{fig:SystemI}, left panel). 
We used the data from Dec 2016 to do the redshift determination, because the seeing was much better.
We also targeted the faint arc to the east, but the signal-to-noise of the data was too low, and we could not obtain a redshift.

We use the \redmapper\ redshift $\zlens=\redshiftRMI$ and the estimated \einsteinradiussub\ of $\thetaesep=\SystemIRadius\arcsec$ to calculate an enclosed mass of $\menclosed=\SystemIMass$ for this system. 


\subsection{\SystemD}
\label{sec:sample:systemD}

\SystemD\ is a cluster-scale lens, where the central lensing galaxy has photometric redshift $\zlens=\redshiftlensD$. 
There is a red arc (\arcA) to the west, a large red arc (\arcB) to the southwest, and a red arc (\arcC) to the south, respectively, as shown in Fig.~\ref{fig:multipanel}d and in the right panel of Fig.~\ref{fig:SystemD}. 
In the follow-up R400 spectroscopy of all three arcs (Fig.~\ref{fig:SystemD}, left panels), we identify an emission line at $\sim7141$\AA, which we identify as \OIIdoublet. 
From this emission line, we obtain redshifts of $\zsrc=\redshiftsrcDA$, $\redshiftsrcDB$, and $\redshiftsrcDC$ (Fig.~\ref{fig:SystemD}). 

The red 'x' near the center of the color image in the upper right panel marks the location of the lens designated for measurement of the Einstein radius.
The position of the red ‘x’ is chosen to simplify the drawing of a circle through the arcs of the source galaxy images.
Because the goal is to obtain a simple estimate of the lens mass, our goals is to first obtain a reasonable estimate of the Einstein radius.

We use the \redmapper\ redshift $\zlens=\redshiftRMD$ and the estimated \einsteinradiussub\ of $\thetaesep=\SystemDRadius\arcsec$ to calculate an enclosed mass of $\menclosed=\SystemDMass$ for this system.

\subsection{\SystemH}
\label{sec:sample:systemH}

\SystemH\ is a galaxy-scale lens with the DES photometric redshift, $\zlens=\redshiftlensH$, and one blue arc (\arcA) to the north. 
The system is shown in Fig.~\ref{fig:multipanel}h and in the right panel of Fig.~\ref{fig:SystemH}. 
Near a wavelength of $\sim9030$\AA, we identify an \OIIdoublet\ emission line in the R400 spectra of both arcs. 
This yields a source redshift of $\zsrc=\redshiftsrcHA$ \arcA\ (Fig.~\ref{fig:SystemH}, left panel). 

We use the \redmagic\ redshift $\zlens = \redshiftRMH$ and the estimated \einsteinradiussub\ of $\thetaesep=\SystemHRadius\arcsec$ to calculate an enclosed mass of $\menclosed=\SystemHMass$ for this system.


\subsection{\SystemE}
\label{sec:sample:systemE}

\SystemE\ is a galaxy-scale system with two small source images, \arcA\ and \arcB, which lie to the southwest of the central red galaxy, as shown in Fig.~\ref{fig:multipanel}e and in the right panel of Fig.~\ref{fig:SystemE}. 
The lensing galaxy has photometric redshift $\zlens=\redshiftlensE$. 
In the R400 spectra for \arcA, we identify emission line features near $\sim6689$\AA, $\sim6942$\AA, $\sim7361$\AA, $\sim7790$\AA, $\sim8723$\AA, $\sim8898$\AA, and $\sim8984$\AA. 
In both images, we take these emission lines to be \OIIdoublet, \NeIII, \Hdelta, \Hgamma, \Hbeta, \OIIIa, \OIIIb, respectively, from which we obtain redshifts of $\zsrc=\redshiftsrcEA$ and $\redshiftsrcEB$ (Fig.~\ref{fig:SystemE}) for these source images. 

We use the \redmagic\ redshift $\zlens=\redshiftRME$ and the estimated \einsteinradiussub\ of $\thetaesep=\SystemERadius\arcsec$ to calculate an enclosed mass of $\menclosed=\SystemEMass$ for this system. 


\subsection{\SystemF}
\label{sec:sample:systemF}

\SystemF\ is a group-scale lens with two central galaxies with DESDM photometric redshifts, $\zlens=\redshiftlensFA$ and $\zlens=\redshiftlensFB$. 
There are two small red arcs, \arcA\ and \arcB, to the east and northeast, respectively, of the central red galaxy. 
These are shown in Fig.~\ref{fig:multipanel}f and in the right panel of Fig.~\ref{fig:SystemF}. 
The R400 spectrum of \arcA\ shows prominent emission lines near five different observed wavelengths---$\sim6385$\AA, $\sim6658$\AA, $\sim9413$\AA, and $\sim10242$\AA, which we identify as \CII, \NeIV, \NeV, and \OIIdoublet, respectively. 
This gives a redshift of $\zsrc=\redshiftsrcFA$. 
The R400 spectrum of \arcB\ presents a similar pattern for these emission lines---$\sim6391$\AA, $\sim6661$\AA, $\sim9413$\AA, and $\sim10242$\AA --- which yields a source redshift of $\zsrc=\redshiftsrcFB$ (Fig.~\ref{fig:SystemF}, left panel).
There is a possible counter-image to the west south-southwest of the lensing galaxy, but it could not be targeted due to its proximity to the central red galaxy and available telescope time. 

The presence of the Ne emission lines in combination with OII lines suggests the possibility that the source is a radio galaxy.  
\citep{humphrey07} identifies NeV and NeIV emission in z$\sim$2.5 radio galaxies as a potential signature of AGN photo-ionization.

The two central galaxies are found in the \redmagic\ catalog with redshifts $\zlens=\redshiftRMFa$ and $\zlens=\redshiftRMFa$. 
We use the \redmagic\ redshifts and the estimated \einsteinradiussub\ of $\thetaesep=\SystemFRadius\arcsec$ to calculate an enclosed mass of $\menclosed=\SystemFMass$ for this system. 


\subsection{\SystemG}
\label{sec:sample:systemG}

\SystemG\ is a group-scale lens. 
The central galaxy has photometric redshift $\zlens=\redshiftlensG$, and there are two blue arcs (\arcA\ and \arcB) to the east and the west. 
These are shown in Fig.~\ref{fig:multipanel}g and in the right panel of Fig.~\ref{fig:SystemG}. 
Near wavelengths of $\sim8920$\AA, we identify \OIIdoublet\ emission lines in the R400 spectra of both arcs. 
These yield source redshifts of $\zsrc=\redshiftsrcGA$ and $\redshiftsrcGB$ for \arcA\ and \arcB, respectively (Fig.~\ref{fig:SystemG}, left panel). 
Only after sky subtraction was the line in \arcB\ revealed. 
Note that \arcB\ appears more extended and much fainter than \arcA, and thus has much lower signal-to-noise. 
The faintness of the source, along with a sky line on top of the data (which affected the subtraction), likely reduced the signal in this emission line. 
The low signal-to-noise for \arcB\ likely contributes to an error in redshift that causes the spectroscopic redshifts to differ beyond their estimated errors.

We use the \redmapper\ redshift $\zlens=\redshiftRMG$ and the estimated \einsteinradiussub\ of $\thetaesep=\SystemGRadius\arcsec$ to calculate an enclosed mass of $\menclosed=\SystemGMass$ for this system. 


\section{Discussion and Summary}\label{sec:summary}

In this paper, we have presented new confirmations of galaxy- to cluster-scale strong lenses in Y1 DES data. 
We first identified candidates in DES data through investigation of cluster sub-samples, and through catalog searches of galaxies based on photometry and proximity of lenses and source images. 
We then visually inspected these subsamples to identify \nbcandidate\ candidates for spectroscopic follow-up.
The search was conducted over 1800 \sqdeg. 
We confirmed these systems with spectroscopy from GMOS on the Gemini South telescope. 
The confirmed sample comprises three galaxy-scale lenses, five group-scale and one cluster-scale lens. 
They have been identified through known emission lines, such as \Lya\ and \OII. 
Of particular note is one system, \SystemF, in which the presence of Ne emission lines suggest it may be a radio galaxy. 
For all the confirmed lenses, we provide a rough estimate of the lens mass based on an \nord{average of the source image-lens separation} from \diehlpaper.
Detailed modeling of these systems can contribute to studies of mass profiles and mass-to-light ratios of early-type galaxies. 
\citet{poh17} will report the modeling of a subset of these confirmed systems. 

The redshift desert is a key problem in the spectroscopic follow-up of strong lenses. 
A number of our candidates could not be confirmed or rejected, because they may exist in a redshift range not covered by GMOS spectrographs. 
One possible solution to this challenge is to seek improved photometric redshift estimations of high-redshift source galaxies.
The measurement or prediction of photometric redshifts of distance objects, is itself  a long-time challenge, largely due to the small number of training sets at high redshift. 
Another solution is to perform observations in a higher wavelength range --- e.g., in the (near) infrared with \nord{Gemini South's FLAMINGOS-2 instrument\footnote{\url{https://www.gemini.edu/sciops/instruments/flamingos2/}}} and Paranal Observatory's Very Large Telescope, which has MUSE and X-Shooter\footnote{\url{https://www.eso.org/public/usa/teles-instr/paranal-observatory/vlt/vlt-instr/} for example}. 

The search in DES Y1 data and in that of \cite{nord16} produced many more candidates than feasibly can be followed up with modern spectroscopic observation resources. 
We confirmed fewer than of those we observed, with an observational strategy that efficiently used the available observing time. 
The best way to improve chances of positive spectroscopic confirmation in future work is to more accurately and precisely predict strong lens candidates from their imaging.  
Future searches of DES data are set to take place with more advanced algorithms, stemming from machine learning and citizen science programs, among others. 
We expect that these algorithms will provide more flexibility, power, and efficiency in identifying high-quality strong lens candidates.



\section*{Acknowledgments}

We are grateful for the extraordinary contributions of our CTIO colleagues and the DES Camera, Commissioning and Science Verification teams for achieving excellent instrument and telescope conditions that have made this work possible. 
The success of this project also relies critically on the expertise and dedication of the DES Data Management organization.

Funding for the DES Projects has been provided by the U.S. Department of Energy, the U.S. National Science Foundation, the Ministry of Science and Education of Spain, the Science and Technology Facilities Council of the United Kingdom, the Higher Education Funding Council for England, the National Center for Supercomputing Applications at the University of Illinois at Urbana-Champaign, the Kavli Institute of Cosmological Physics at the University of Chicago, the Center for Cosmology and Astro-Particle Physics at the Ohio State University, the Mitchell Institute for Fundamental Physics and Astronomy at Texas A\&M University, Financiadora de Estudos e Projetos, Funda\c{c}\~{a}o Carlos Chagas Filho de Amparo \`{a} Pesquisa do Estado do Rio de Janeiro, Conselho Nacional de Desenvolvimento Científico e Tecnol\'{o}gico and the Minist\'{e}rio da Ci\^{e}ncia e Tecnologia, the Deutsche Forschungsgemeinschaft and the Collaborating Institutions in the Dark Energy Survey. The DES data management system is supported by the National Science Foundation under Grant Number AST-1138766. 
The DES participants from Spanish institutions are partially supported by MINECO under grants AYA2012-39559, ESP2013-48274, FPA2013-47986, and Centro de Excelencia Severo Ochoa SEV-2012-0234, some of which include ERDF funds from the European Union.

The Collaborating Institutions are Argonne National Laboratory, the University of California at Santa Cruz, the University of Cambridge, Centro de Investigaciones Energ\'{e}ticas, Medioambientales y Tecnol\'{o}gicas-Madrid, the University of Chicago, University College London, the DES-Brazil Consortium, the Eidgenoessische Technische Hochschule (ETH) Zurich, Fermi National Accelerator Laboratory, the University of Edinburgh, the University of Illinois at Urbana-Champaign, the Institut de Ci\`{e}ncies de l'Espai (IEEC/CSIC), the Institut de F\'{i}sica d'Altes Energies, Lawrence Berkeley National Laboratory, the Ludwig-Maximilians Universit\"{a}t and the associated Excellence Cluster Universe, the University of Michigan, the National Optical Astronomy Observatory, the University of Nottingham, the Ohio State University, the University of Pennsylvania, the University of Portsmouth, SLAC National Accelerator Laboratory, Stanford University, the University of Sussex, and Texas A\&M University.

This work is based in part on observations obtained at the Gemini Observatory, which is operated by the Association of Universities for Research in Astronomy, Inc., under a cooperative agreement with the NSF on behalf of the Gemini partnership: the National Science Foundation (United States), the National Research Council (Canada), CONICYT (Chile), the Australian Research Council (Australia), Minist\'{e}rio da Ci\^{e}ncia, Tecnologia e Inova\c{c}\~{a}o (Brazil) and Ministerio de Ciencia, Tecnolog\'{i}a e Innovaci\'{o}n Productiva (Argentina). The data was processed using the Gemini IRAF package v2.16.

This research has made use of NASA's Astrophysics Data System.

This manuscript has been authored by Fermi Research Alliance, LLC under Contract No. DE-AC02-07CH11359 with the U.S. Department of Energy, Office of Science, Office of High Energy Physics. The United States Government retains and the publisher, by accepting the article for publication, acknowledges that the United States Government retains a non-exclusive, paid-up, irrevocable, world-wide license to publish or reproduce the published form of this manuscript, or allow others to do so, for United States Government purposes.

\bibliography{DESStrongLensY1}
\bibliographystyle{apj}


\tablelensobjects

\tablelensobjectsnotconfirmed

\tablerankobserve

\tableobservationlog

\tablelensingfeatures


\figuremultipanel 

\FigureSystemA
\FigureSystemB
\FigureSystemC
\FigureSystemI
\FigureSystemD
\FigureSystemH
\FiguresystemE
\FigureSystemF
\FigureSystemG



\end{document}